\documentclass[graybox, nosecnum]{svmult}

\usepackage{mathptmx}       
\usepackage{helvet}         
\usepackage{courier}        
\usepackage{type1cm}        
\usepackage{makeidx}         
\usepackage{graphicx}        
\usepackage{multicol}        
\usepackage[bottom]{footmisc}
\usepackage{hyperref}        
\usepackage{soul}            
\hypersetup{colorlinks=true,urlcolor=blue}
\usepackage[square,numbers]{natbib}
\makeindex             

\usepackage{xspace}       %

\newcommand{\Al}{$^{26}${Al}\xspace}
\newcommand{\about}{$\simeq$}

\newcommand{\Fe}{$^{60}$Fe\xspace}

\newcommand{\Ni}{$^{56}$Ni\xspace}
\newcommand{\Ti}{$^{44}$Ti\xspace}

\newcommand{\Msol}{M\ensuremath{_\odot}\xspace}

\newcommand{\araa}{Ann.Rev.Astron.\&Astroph. }%
\newcommand{\apj}{Astrophys. J. }%
\newcommand{\apjl}{Astrophys. J. Lett. }%
\newcommand{\apjs}{Astrophys. J. Suppl. Ser. }%
\newcommand{\aap}{Astron. Astrophys. }%
\newcommand{\aapr}{Astron. Astrophys.~Rev. }%
\newcommand{\aaps}{Astron. Astrophys. Suppl. }%
\newcommand{\aj}{Astron. J. }%
\newcommand{\mnras}{Mon. Notices Royal Astron. Soc. }%
\newcommand{\nar}{New Astron. Rev. }%
\newcommand{\prc}{Phys.~Rev.~C }%
\newcommand{\pasa}{PASA }%
\newcommand{\solphys}{Sol.~Phys. }%
\newcommand{\ssr}{Space~Sci.~Rev. }%
\newcommand{\nat}{Nature }%
\newcommand{\physrep}{Phys.~Rep. }%
\makeindex             

\begin{document}

\title*{Radioactive decay}
\author{Roland Diehl}
\institute{Roland Diehl \at Max Planck Institut f\"ur extraterrestrische Physik, 85748 Garching, Germany, \email{rod@mpe.mpg.de}}
%
%
\maketitle

\abstract{Radioactive decay of unstable atomic nuclei leads to liberation of nuclear binding energy in the forms of gamma-ray photons and secondary particles (electrons, positrons); their energy then energises surrounding matter. Unstable nuclei are formed in nuclear reactions, which can occur either in hot and dense extremes of stellar interiors or explosions, or from cosmic-ray collisions. In high-energy astronomy, direct observations of characteristic gamma-ray lines from the decay of radioactive isotopes are important tools to study the process of cosmic nucleosynthesis and its sources, as well as tracing the flows of ejecta from such sources of nucleosynthesis. These observations provide a valuable complement to indirect observations of radioactive energy deposits, such as the measurement of supernova light in the optical.
Here we present basics of radioactive decay in astrophysical context, and how gamma-ray lines reveal details about stellar interiors, about explosions on stellar surfaces or of entire stars, and about the interstellar-medium processes that direct the flow and cooling of nucleosynthesis ashes once having left their sources. We address radioisotopes such as $^{56}$Ni, $^{44}$Ti, $^{26}$Al, $^{60}$Fe, $^{22}$Na, $^{7}$Be, and also how characteristic gamma-ray emission from the annihilation of positrons is connected to these.}

\section{Basics of Radioactivity}
\subsection{Discovery}
In the nineteenth century, various efforts aimed to bring order into the \emph{elements} encountered in nature. The inventory of the elements  assembled by the Russian chemist Dimitri Mendeleyev \index{Mendeleyev, D.} in 1869 grouped elements according to their chemical properties as derived from the compounds they were able to form, at the same time sorting the elements by atomic weight. The ordering by Mendeleyev enforced gaps in the table, for expected but then unknown elements. 
In the second half of the nineteenth century scientists were all-excited about chemistry and the fascinating discoveries around elements with their properties and their apparent transformations in chemical reactions. Today the existence of 118 elements is firmly established\footnote{IUPAC, the international union of chemistry, coordinates definitions, groupings, and naming; see www.IUPAC.org}. Element 118, called oganesson (Og), is the most massive superheavy element which has been synthesised, and found to exist at least for short time intervals. More massive elements may still exist in an island of stability beyond.  The latest additions, no. 113-118, all were discovered in the year 2016, which reflects the concerted experimental efforts.

After Conrad R\"ontgen's  discovery in 1895 of {\it X-rays} as a type of penetrating electromagnetic radiation,  Antoine Henri Becquerel  discovered radioactivity in 1896, as he was engaged in chemical experiments in his research of photographic-plate materials regarding phosphorescence. At the time, Becquerel had prepared some plates treated with uranium-carrying minerals, but did not get around to make the planned experiment. When he found the plates in their dark storage some time later, he accidentally processed them, and was surprised to find an image of a coin which happened to have been stored with the plates. Excited about X-rays, he believed he had found yet another type of penetrating  radiation. 
Within a few years, Becquerel with Marie and Pierre Curie  and others recognised that the origin of the observed radiation were elemental transformations of the uranium minerals: The physical process of {\it radioactivity} had been found! But when sub-atomic particles and the atom were discovered at the beginning of the twentieth century, the revolutionary aspect of elements being able to spontaneously change their nature was drowned by such new excitement. Still, well before atomic and quantum physics began to unfold, the physics of {\it weak interactions} had already been discovered, in its form of {\it radioactivity}.

\begin{figure} 
  \includegraphics[width=\textwidth]{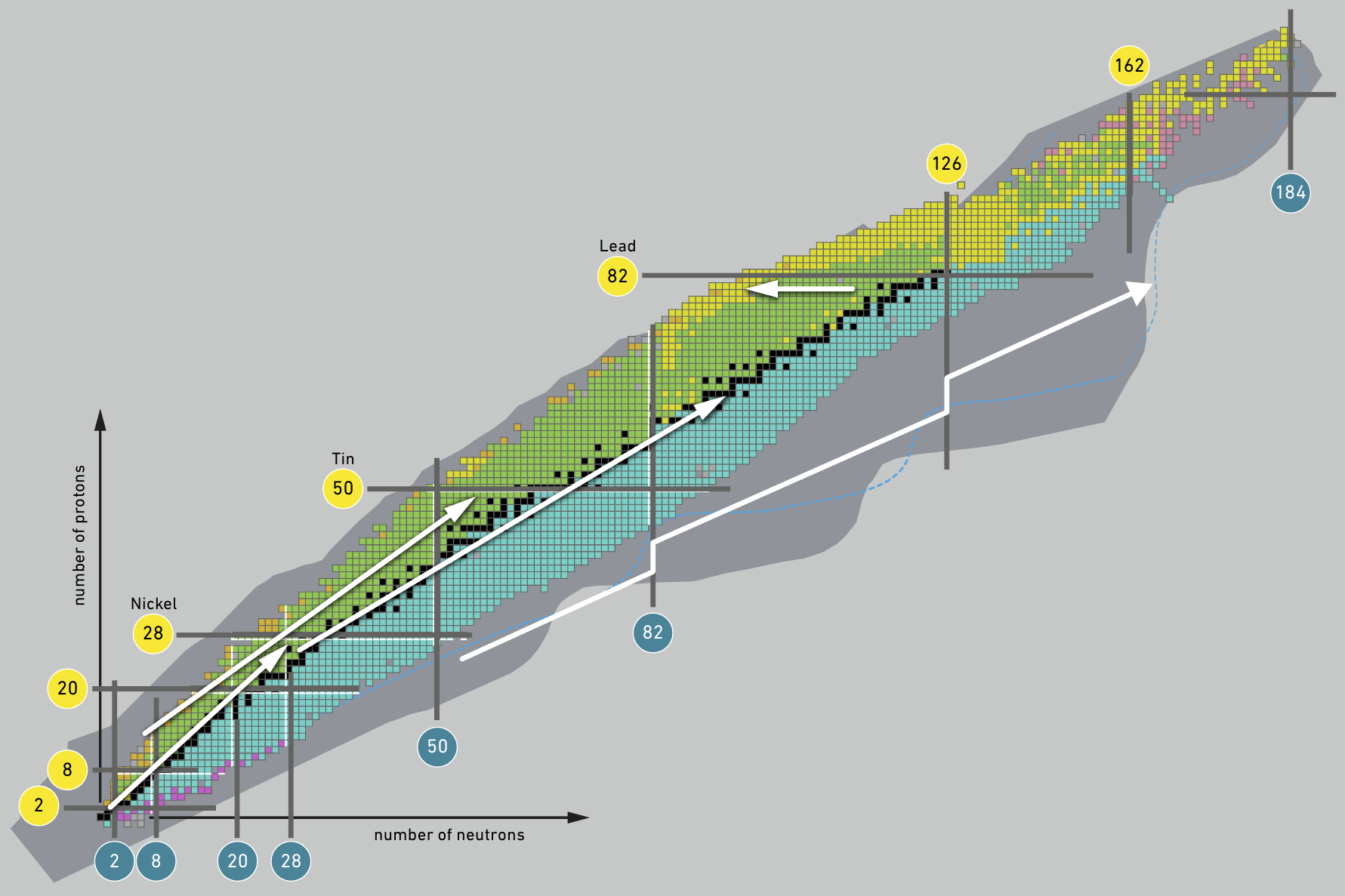}
  \caption{The table of isotopes, showing the possible configurations of atomic nuclei in a chart of neutron number (abscissa) versus proton number (ordinate). Numbers identify the \emph{magic numbers} of nucleons, which characterise the most tightly-bound configurations. White arrows indicate the nuclear-reaction paths of specific \emph{ processes of nucleosynthesis} in cosmic sites. The stable isotopes are marked in black.
  All other isotopes (in color) are unstable, or radioactive; they will decay until a stable nucleus is obtained. }
  \label{fig:table-of-isotopes}
\end{figure} 

The atomic nucleus is composed of \emph{hadrons}, the protons and neutrons, which are subject to the strong nuclear force, which so binds an atomic nucleus.
The landscape of nuclear configurations is illustrated in Fig.~\ref{fig:table-of-isotopes}, showing numbers of neutrons as abscissa and number of protons as ordinate, with black symbols as the naturally-existing stable isotopes, and coloured symbols for unstable isotopes. The latter are subject to radioactive decay towards stable isotopes, with less total binding energy per nucleon.

Often, the result of such inner transformations of the hadronic configuration produces a daughter nucleus in an excited state. The transition to the ground state then involves spin changes, emitting photons to carry away spin and energy. These are the characteristic photons that accompany most radioactive decays. 
 
The production of non-natural isotopes and thus the generation of man-made radioactivity led to the Nobel Prize in Chemistry for Jean Fr\'ed\'eric Joliot-Curie and his wife Ir\'ene in 1935 -- the second Nobel Prize awarded for the subject of radioactivity after the 1903 prize went jointly to Pierre Curie, Marie Sk\l odowska Curie, and Henri Becquerel, also in the field of Chemistry.

\subsection{Characteristics} 
The probability per unit time for a single radioactive nucleus to decay is independent of the age of that nucleus. Unlike our common-sense experience with living things, or with astrophysical objects that evolve, such as stars, radioactive decay does not become more likely as the nucleus ages. 
In \emph{$\beta$-decays}, the transition is mediated by the \emph{weak  interaction}, transforming a  neutron into a proton and vice versa.
 The mass difference of the neutron and the proton (plus an electron for charge neutrality)  is \citep{Patrignani:2016} 1.293332 MeV = 939.565413 - 938.272081 MeV for the mass of neutron and proton, respectively. Neutrons are unstable and decay from the weak interaction, with a mean life of 880 seconds, into a proton, an electron, and an anti-neutrino, releasing this mass difference in kinetic energy. This is the origin of radioactivity. 
When neutrons and protons transform through such weak interaction, the new configuration is unstable; the atomic nucleus can (and must) find a new stable configuration of its hadrons, lowering the total nuclear binding energy. The excess binding energy thus can be released. Depending on the direction of the transformation $p\longleftrightarrow n$, $\beta^-$ or $\beta^+$~decays are characterised by emission of an electron, or positron, respectively, for charge conservation.
\begin{equation}\label{eq_n-decay}
n \longrightarrow p \mbox{ } + e^- \mbox{ } + \overline{\nu_e} 
\end{equation}
\begin{equation}\label{eq_p-decay}
p \longrightarrow n \mbox{ } + e^+ \mbox{ } + {\nu_e}
\end{equation}
If such a transformation occurs inside an atomic nucleus, the quantum state of the nucleus as a whole is altered. Depending on the variety of configurations in which this new state may be realised (i.e. the \emph{phase space} available to the decaying nucleus), this transformation may be more or less likely, as the total energy of a composite system of nucleons aims at a minimum value. 

\noindent {\it Beta decay} is the most-peculiar radioactive decay type, as it is caused by the nuclear \emph{weak interaction} which converts neutrons into protons and vice versa. The neutrino $\nu$ carries energy and momentum to balance the dynamic quantities. 

There are three types\footnote{We ignore here two additional $\beta$ decays which are possible from $\nu$ and $\overline{\nu}$ captures, due to their small probabilities.} of $\beta$-decay: 
\begin{equation}\label{eq_beta+-decay}
^A_ZX_N\mbox{ }  \longrightarrow \mbox{ } ^A_{Z-1}X_{N+1} \mbox{ } + e^+ \mbox{ } + \nu_e
\end{equation}
\begin{equation}\label{eq_beta--decay}
^A_ZX_N \mbox{ } \longrightarrow \mbox{ } ^A_{Z+1}X_{N-1} \mbox{ } + e^- \mbox{ } + \overline{\nu_e}
\end{equation}
\begin{equation}\label{eq_beta--decay}
^A_ZX_N \mbox{ } + e^- \longrightarrow \mbox{ } ^A_{Z-1}X_{N+1}  \mbox{ } + {\nu_e}
\end{equation}
In addition to \emph{$\beta^-$~decay}, these are the conversion of a proton into a neutron (\emph{$\beta^+$~decay}), and \emph{electron capture}.
The weak interaction itself involves two different aspects with intrinsic and different strength, the  vector and axial-vector couplings. 
These result in  \emph{Fermi} and \emph{Gamow-Teller transitions}, respectively \citep[see][for a review of weak-interaction physics in nuclear astrophysics]{Langanke:2003a}.

An example of $\beta$ decay is \index{isotopes!13N} $_7^{13}$N $\longrightarrow \mbox{ }_6^{13}$C~+~e$^{+}$~$+$~$\nu$, having mean lifetime $\tau$ near 10 minutes. This contributes to the early light from nova explosions. The kinetic energy $Q$ of the two leptons, as well as the created electron's mass, must be provided by the radioactive nucleus having greater mass than the sum of the masses of the daughter nucleus and of an electron (neglecting the comparatively-small neutrino mass).
\begin{equation}
Q_{\beta} =[M(_7^{13}\rm{N}) - M(_6^{13}\rm{C})- m_{e}]c^2
\end{equation}
\noindent where these masses are nuclear masses, not atomic masses. A small fraction of the energy release $Q_{\beta}$ appears as the recoil kinetic energy of the daughter nucleus, but the remainder appears as the kinetic energy of electron and of neutrino.

Capture of an electron is a \emph{two-particle} reaction, the bound atomic electron $e^{-}$ or a free electron in hot plasma being required for this type of $\beta$ decay. Therefore, depending on availability of the electron,  lifetimes in electron-capture $\beta$ decay can be very different for different environments. 
	In the laboratory case, electron capture usually involves the $1s$ electrons of the atomic structure surrounding the radioactive nucleus, because those hold their largest density at the nucleus. 
		The situation for electron capture processes differs significantly in the interiors of stars and supernovae: Nuclei are fully-ionized in plasma at such high temperature. The capture lifetime of $^7$Be, for example, which is 53 days against $1s$ electron capture in the laboratory, is lengthened to about 4 months in the central region of our Sun.

	The range of the $\beta$ particle (its \emph{stopping length}) in normal terrestrial materials is small, being a charged particle which undergoes Coulomb scattering. An MeV electron has a range of several meters in standard air, during which it loses energy by ionisations and inelastic scattering. In tenuous cosmic plasma such as in supernova remnants, or in interstellar gas, such collisions, however, become rare, and may be unimportant compared to electromagnetic interactions of the magnetic field (\emph{collisionless plasma}).
Energy deposit or escape is a major issue in intermediate cases, such as in the expanding envelopes of stellar explosions, in supernovae (positrons from $^{56}$Co  and $^{44}$Ti) and in novae (many $\beta^+$ decays such as $^{13}$N). 
Even in small solids and dust grains, energy deposition from \Al  $\beta$-decay, for example, injects 0.355~W~kg$^{-1}$ of heat. This is sufficient to result in melting signatures, which have been used to study condensation sequences of solids in the early solar system, and are believed to control the water content in newly-forming planets \citep{Lichtenberg:2019}.

\noindent {\it Gamma decay} is another expression for the de-excitation of a nucleus from its excited configuration of the nucleons to a lower-lying state of the same nucleons. We denote such electromagnetic transitions of an excited nucleus \emph{radioactive $\gamma$-decay} when the decay time of the excited nucleus is considerably longer than what is typical for excited nuclei, and that nucleus thus may be considered a temporarily-stable configuration of its own, a \emph{metastable} nucleus.  Typically, a nucleus relaxes  in an intrinsically-fast process, and lifetimes for excited states of an atomic nucleus are 10$^{-9}$seconds. 

The spin (angular momentum) is a conserved quantity of the system in such a transition. The spin of a nuclear state is a property of the nucleus as a whole, and reflects how the states of protons and neutrons are distributed over the quantum-mechanically allowed \emph{shells} or nucleon wave functions (as expressed in the \emph{shell model} view of an atomic nucleus).
The emitted photon (\emph{$\gamma$ ray}) carries a \emph{multipolarity} that results from the spin differences of initial and final states of the nucleus. Dipole radiation is most common and has multipolarity 1, emitted when initial and final state have angular momentum difference $\Delta l=1$. Quadrupole radiation (multipolarity 2, from $\Delta l=2$) is $\sim$6 orders of magnitude more difficult to obtain, and likewise, higher multipolarity transitions are becoming less likely as probability decreases in this way. 
This explains why some excited states in atomic nuclei are much more long-lived (\emph{meta-stable}) than others; their transitions to the ground state are also considered as \emph{radioactivity}, and called \emph{$\gamma$ decay}.

The range of a $\gamma$-ray (its \emph{stopping length}) is typically about 5-10~g~cm$^{-2}$ in passing through matter of all types. Hence, except for dense stars and their explosions, radioactive energy from $\gamma$ decay is of astronomical implication only.

In radioactive decay, the number of decays at each time is proportional to the number of currently-existing radioisotopes:  
\begin{equation}
 {{dN}\over{dt}} =  -\lambda \cdot N 
\end{equation}
Here $N$ is the number of isotopes, and the {\it radioactive-decay constant} $\lambda$ is the characteristic of a particular radioactive species.

Therefore, in an ensemble consisting of a large number of identical and unstable isotopes, their number remaining after radioactive decay  declines exponentially with time:
\begin{equation}
\label{eq_1} 
 N = N_0 \cdot e^{-{t\over\tau}} 
\end{equation}
The decay time $\tau$ is the inverse of the radioactive-decay constant, and  $\tau$ characterises the time after which the number of isotopes is reduced by decay to $1/e$ of the original number. Correspondingly, the radioactive half-life $T_{1/2}$,  is defined as the time after which the number of isotopes is reduced by decay to $1/2$ of the original amount, with
\begin{equation}
   T_{1/2} = {\tau \over ln(2)} 
\end{equation}

In the general laboratory situation, radioactive decay involves a transition from the ground state of the parent nucleus to the daughter nucleus in an excited state. But in cosmic environments, nuclei may be part of hot plasma, and temperatures exceeding millions of degrees lead to excited states of nuclei being populated. Thus, quantum mechanical transition rules may allow and even prefer other initial and final states, and the nuclear reactions involving a radioactive decay become more complex. Excess binding energy will be transferred to the end products, which are the daughter nucleus plus emitted (or absorbed, in the case of electron capture transitions) leptons (electrons, positrons, neutrinos) and $\gamma$-ray photons. 

In cosmic hot plasma, the occupancy of nuclear states may also be affected by the \emph{thermal} excitation spectrum of the \emph{Boltzmann distribution} of particles, populating states at different energies according to:
\begin{equation}
{dN \over dE} = G_j \cdot e^{-{{E}\over{k_BT}}} 
\end{equation}
\noindent Here $k_B$ is Boltzmann's constant, $T$ the temperature of the particle population, $E$ the energy, and $G_j$ the statistical weight factor of all different possible states $j$ which correspond to a specific energy\footnote{States may differ in their quantum numbers, such as spin, or orbital-momenta projections; if they obtain the same energy $E$, they are called \emph{degenerate}.} $E$.
Inside stars, and more so in explosive environments, temperatures can reach ranges which are typical for nuclear energy-level differences. Therefore, in cosmic sites, radioactive decay time scales may be significantly different from what we measure in terrestrial laboratories on \emph{cold} nuclei.

Also the atomic-shell environment of a nucleus may modify radioactive decay,  if a decay involves \emph{capture or emission of an electron}  to transform a proton into a neutron, or vice versa. Electron capture decays are inhibited in fully-ionized plasma, due to the non-availability of electrons. Also $\beta^-$-decays are affected, as the phase space for electrons close to the nucleus is influenced by the population of electron states in the atomic shell.

\begin{figure}
\centerline{
  \includegraphics[width=0.6\textwidth]{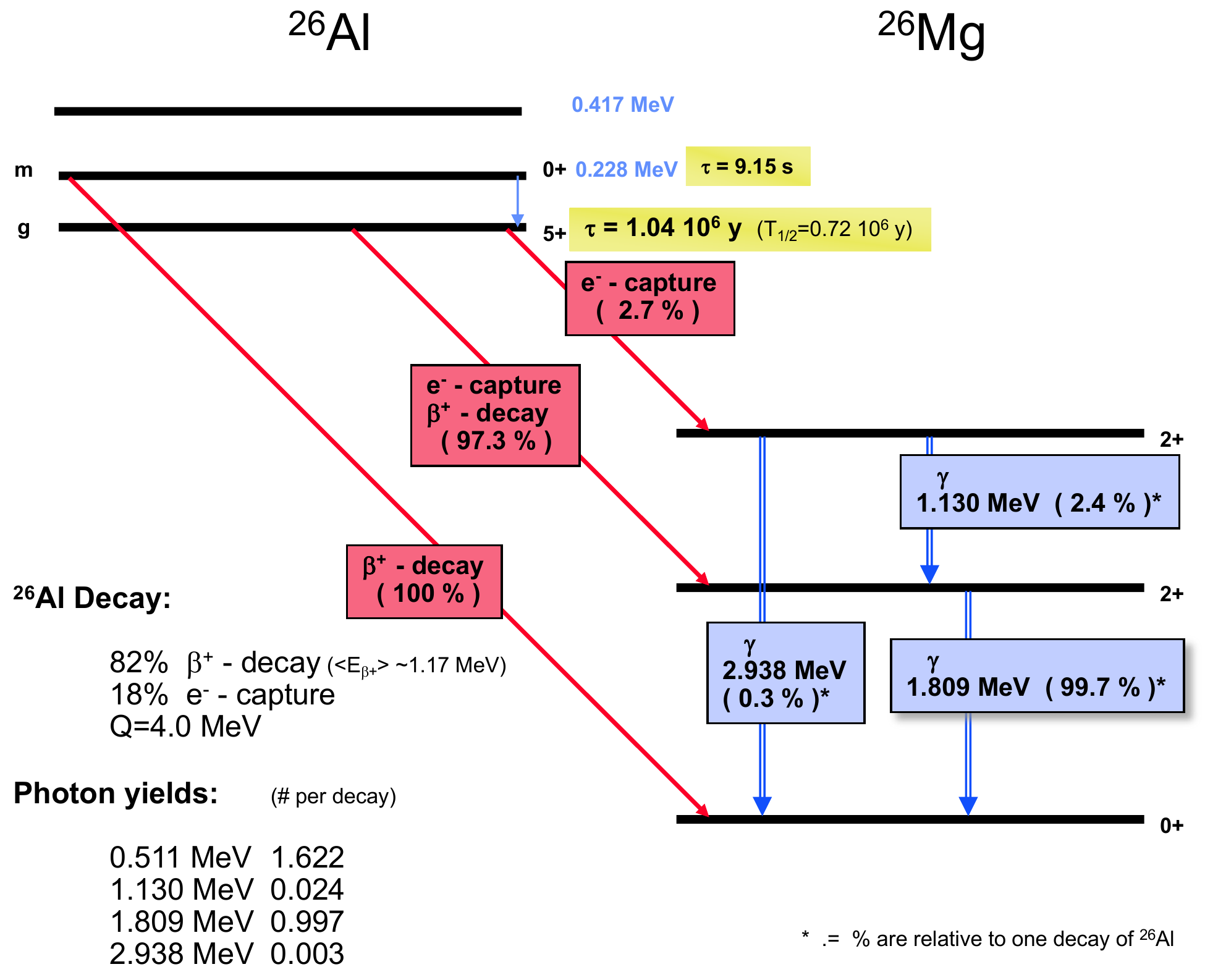}}
  \caption{\Al decay, with different states of the $^{26}$Al and $^{26}$Mg nuclei and the possible transitions between. Both electron capture and $\beta^+$ decay occur, and the latter leads to positron annihilation $\gamma$~rays, in addition to the de-excitation $\gamma$~rays from transitions within the $^{26}$Mg nucleus. Note that decay from the first excited state of $^{26}$Al does not lead to $\gamma$-ray emission (see text for details). }
  \label{fig:26Al-decay}
\end{figure}

An illustrative example of radioactive decay is the \Al nucleus, illustrated in Fig.~\ref{fig:26Al-decay}. The ground state of \Al is a $5^+$ state. Lower-lying states of the neighbouring isotope $^{26}$Mg have states $2^+$ and $0^+$, so that a rather large change of angular momentum $\Delta l$ must be carried by radioactive-decay secondaries. This explains the large $\beta$-decay lifetime of \Al of $\tau\sim$1.04$\times$10$^6$~y.  
In the level scheme of \Al, excited states exist at energies 228, 417, and 1058~keV. The $0^+$ and $3^+$ states of these next excited states are more favourable for decay due to their smaller angular momentum differences to the $^{26}$Mg states, although $\Delta l=0$ would not be \emph{allowed} for the 228~keV state to decay to $^{26}$Mg's ground state. This explains its relatively long lifetime of 9.15~s, and it is called a \emph{metastable} state of \Al.  If thermally excited, which would occur in nucleosynthesis sites exceeding a few 10$^8$K, \Al may decay through this state without $\gamma$-ray emission, while the ground state decay is predominantly a \emph{$\beta^+$ decay} through excited $^{26}$Mg states and thus including $\gamma$-ray emission. Secondary products, lifetime, and radioactive energy available for deposits and observation depend on the environment. 


\section{Radioactivity in Astrophysics} 

\subsection{General considerations}
Nuclear reactions in cosmic sites re-arrange the basic constituents of atomic nuclei (neutrons and protons) among the different allowed configurations.
Radioactive, i.e. unstable, isotopes are a result of such nuclear reactions, as a by-product. The existence of radioactive isotope reflects the previous occurrence of those nuclear reactions. The radioactive decay of isotopes provides energy input and compositional changes, leading to observable consequences. Hence, observations of radioactive decay provide a way to learn about astrophysical processes. 
 
\begin{figure}
  \centering
  \includegraphics[width=\columnwidth]{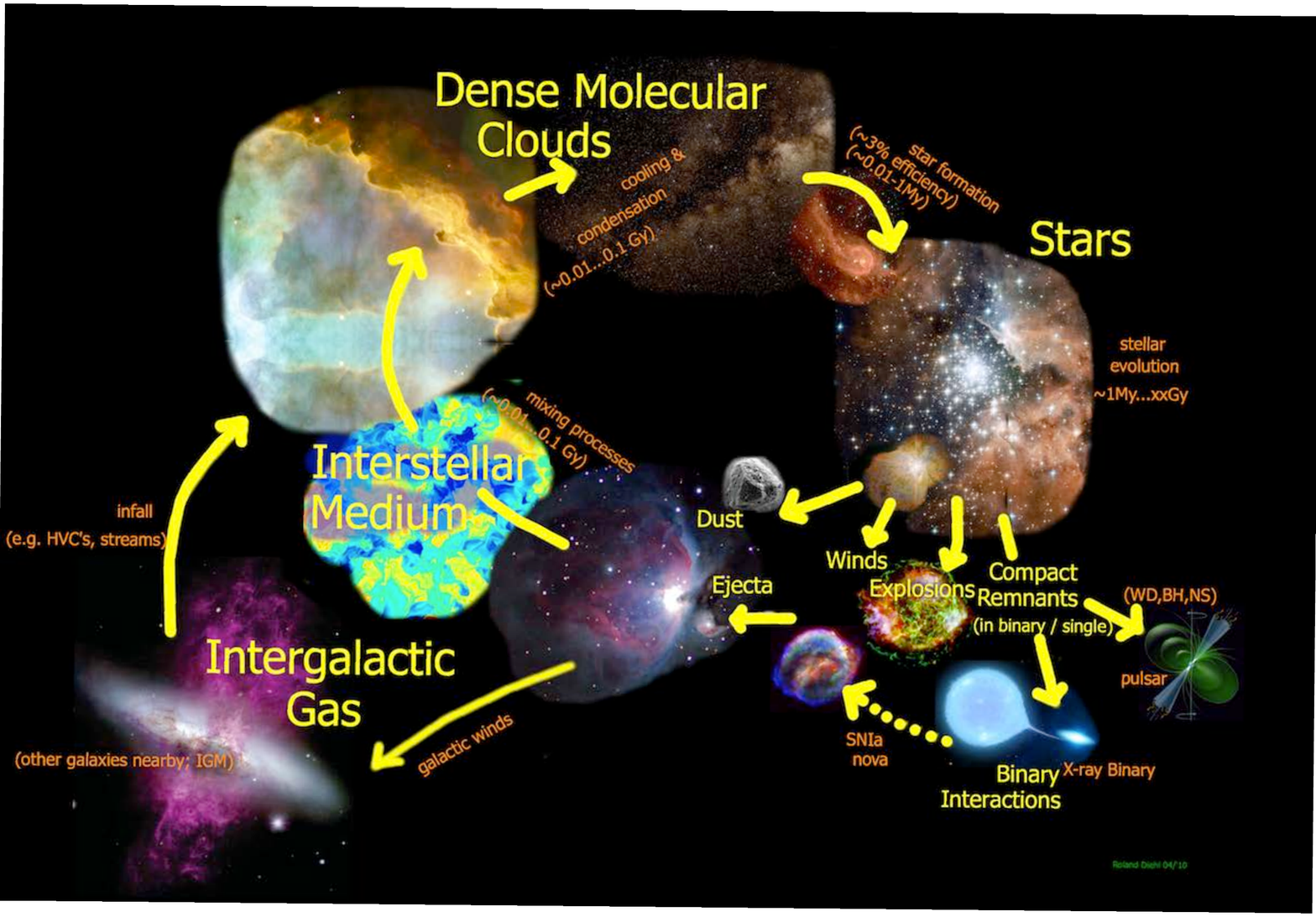}
   \caption{The cycle of matter, with stars forming out of interstellar gas, evolving towards wind and explosions releasing newly-formed isotopes, that are then fed back into the next generation of stars.}
  \label{fig_matter-cycle}
\end{figure}

The composition of cosmic material in the current universe and its observable objects is the result of nuclear reactions throughout cosmic history from its beginnings to the isolation of an object from further nuclear processing. 
Big Bang Nucleosynthesis about 13.8~Gyrs ago  left behind a primordial composition where hydrogen (protons) and helium were the most-abundant species; the total amount of nuclei heavier than He (the \emph{metals}) was less than 10$^{-9}$ (by number, relative to hydrogen) \citep{Cyburt:2016}. 
Today, the total mass fraction of metals in matter with \emph{solar abundances} is\footnote{Our local reference for cosmic material composition seems to be remarkably universal, and representative for the local universe. Note that the Sun formed 4.6~Gy ago, hence this composition sample is from a time where the Galaxy was little more than half its current age. Solar composition is still debated: earlier than $\sim$2005, the commonly-used value for solar metallicity \index{metallicity!solar} had been 2\%. } $Z=0.0134$  \citep{Asplund:2009}, i.e. of order $\sim$percent, compared to a hydrogen mass fraction of $X=0.7381$. 
This growth of metal abundances by about seven orders of magnitude is the effect of cosmic nucleosynthesis. Nuclear reactions in stars, supernovae, novae, and other places where nuclear reactions may occur, all contribute. But it also is essential that at least a fraction of the nuclear-reaction products inside those cosmic objects will eventually be made available to other cosmic gas and solids, and thus to later-generation stars such as our solar system born 4.6~Gyrs ago. 
This cycling of material is illustrated in Figure~\ref{fig_matter-cycle}, and includes radioactive contributions in the ejecta from stellar nucleosynthesis.

Throughout cosmic evolution, nuclear reactions occur in various sites with different characteristic environmental properties. Each reaction environment leads to rearrangements of the relative abundances of cosmic nuclei. Winds, explosions, and binary mass transfers can liberate some of those reaction products from the compact centers of stars and their explosions, while, however, a major fraction of reaction products is buried in compact remnants of stellar evolution, such as white dwarfs, neutron stars, and black holes. 
For some of those compact white dwarfs and neutron stars, interactions with a companion star within a binary system can lead some time later to ejections of material again, e.g. in the forms of novae or thermonuclear supernovae. 
The cumulative process of nuclear transformations and return of some of the products into the interstellar and stellar materials is called \emph{cosmic chemical evolution}.
Radioactivity is among the most-directly related processes that can tell us about the astrophysical processes during cosmic chemical evolution.

The phenomenon of radioactivity has impacts on astrophysical investigations in two fundamental aspects:
\begin{enumerate}
\item The presence of radioactive isotopes changes processes in astrophysical sites and objects from what is known from laboratories .
\item The presence of isotopes results in phenomena, the observations of which enables new and characteristic lessons on astrophysics.
\end{enumerate}

\subsubsection*{Different Processes}
The existence of radioactive isotopes implies that high-energy processes which exceed the threshold of production for such isotopes will produce such output. Therefore, radioactivity presents a channel for absorbing energy in high-energy collisions, through incorporation of kinetic/external energy of the collision into internal nuclear binding or excitation energy. The nuclear reactions in nucleosynthesis environments that transform combinations of nucleons into others can be considered to mediate the thermal energy reservoir of a system with the cumulative nuclear binding energy that is represented by a particular composition of atomic nuclei. (1a)

The presence of radioactive material implies that the composition of that cosmic material will change over time, due to radioactive decay. Thus, for all physical and chemical processes which depend on composition, there will be a time-dependent component in such process. (1b)

The presence of radioactive material implies that radioactive decay will liberate nuclear binding energy, in the forms of $\gamma$-radiation and of energized daughter products. Both of these contribute to the heat of the respective environment, and thus to its thermal luminosity, as much as this energy is captured or absorbed by such environment. (1c)

Depending on the astrophysical objective, radioactive isotopes may be called \emph{short-lived}, or \emph{long-lived}, depending on how the radioactive lifetime compares to astrophysical time scales of interest. Examples are the utilisation of \Al and \Fe ($\tau\sim$My) to trace cumulative nucleosynthesis over a time interval of several million years (\emph{long-lived}), or of \Ni and \Ti to trace how supernovae explode  (\emph{short-lived}). Note that in cosmic chemical evolution, on the other hand, \Al and \Fe would be called  \emph{short-lived}, because radioactive isotopes such as from Th and U with decay times of Gyrs are used for temporal evolution studies on cosmological time scales; \Al is also studied from meteorites with respect to the early solar system, and measures the nucleosynthesis activity near the presolar nebula 4.6 Gyrs ago within a precision of Myrs.

\subsubsection*{New Astronomy} 

Direct astronomical measurements of radioactivity use two main methods: Characteristic nuclear emission lines measured with gamma-ray telescopes, and isotopic abundances in samples of cosmic matter captured within our solar system.  
Both of these methods are rather new and not-so familiar disciplines of astronomy. 
 They are complemented by less-direct measurements of radioactivity, reflecting the differences in processes due to radioactivity as discussed above.

The detection of the presence of radioactive nuclei in specific cosmic samples of material can be used to trace the past history and evolution of such sample: The amount of radioactivity reflects the time and intensity of energetic interactions, or exposure to a source of high energy; its characteristic drop in intensity verifies that this particular messenger is the source of information.
Examples are: the presence of Tc in spectra of giant stars, which is the astronomical proof of recent nuclear reactions within this star \citep{Merrill:1952}; then the anomalous isotopic ratio of Ne found in meteorites after they were heated to above 1000~K \citep[see][for more details of this route to the discovery of stardust]{Clayton:2018}; and the observed composition of cosmic rays near Earth which includes radio-isotopes resulting from interstellar spallation reactions, such as $^{10}$Be, $^{36}$Cl, $^{26}$Al \citep{Mewaldt:2001}. 

A changing composition due to radioactive decay is best observed through the abundance of a daughter isotope which exceeds the isotope abundance ratios for the respective element. For example, excess $^{26}$Mg in Al-rich inclusions of meteorites is interpreted as a result of decay of $^{26}$Al, and thus implies that an an earlier time, radioactive $^{26}$Al had been present in such material. This is the origin of a hypothesis that the early solar system had been enriched in some unexpected way with material carrying this radioactive isotope. 

Such changes in isotopic abundance ratios also provide \emph{a cosmic clock}: From an initial abundance of radioactive material, decay enriches the daughter isotope and depletes the parent isotope, strictly following the exponential law of radioactive decay (see above). So, once an initial isotope abundance ratio is known, and material is isolated from any other influx or depletion of the specific isotopes, the time since isolation of such material can be calculated from the characteristic radioactive decay time that can be obtained from nuclear experiment or theory. We know $^{14}$C dating from civilian life, e.g. conserving an atmospheric ratio $^{14}$C/$^{12}$C of typically  1.2$\times$10$^{-12}$ in plants, which can be age-dated using the $^{14}$C half life of 5700 years, and counting the remaining  $^{14}$C and $^{12}$C abundances with an AMS machine and single-ion detectors or current counters, respectively \citep{Kutschera:2013}. 
Section~1 of the following Chapter 
below discusses specific astrophysical applications using tools of high-energy electromagnetic radiation as covered in this Handbook.

Nuclear reactions in the inner regions of stars release nuclear binding energy, some of this directly as part of the fusion reaction, but mostly through the fusion products and the radioactive processes therein, as discussed above. This energy release counterbalances the gravitational contraction, and thus can stabilise the star, as long as nuclear energy release is adjusted to gravitational pull and the energy loss due to radiation from the surface. Therefore, nuclear energy release stabilised the star as a long-lived object, and makes it shine, as it processes a \emph{fuel} of lighter to more-tightly bound nuclei \citep{Eddington:1919}. 
Exhaustion of a fuel terminates this inner energy source, compression ensues and enables a next nuclear energy release process.   
Successive burning stages during the evolution of a star continue releasing nuclear energy and producing radioactivity; at each stage, they correspond to different nuclear fuels. At each stage, this energy release temporarily slows down the gravitational collapse of the star, with contraction and compressional heating in short transitional phases. Radiation transfer from $\gamma$-ray energies to thermal energies occur in the large stellar envelope, so that the thermal emission of starlight from the surface has lost all information about the origin of the energy source and its shaping by radioactivity.

In supernova explosions, the strive for most-tightly bound nucleons near nuclear statistical equilibrium leads to a production of large amounts of radioactive $^{56}$Ni \citep{Arnett:1996a}: Of order 0.1~\Msol are typically produced in core-collapse supernovae, as we know from SN1987A (0.07~\Msol)\citep{Arnett:1989a,Fransson:2002,McCray:2016}. Thermonuclear supernovae (type Ia) produce even more, typically 0.5~\Msol\ \citep{Scalzo:2014,Seitenzahl:2017}.
SN1987A was the first core-collapse supernova where gamma rays directly originating from the radioactive decay of $^{56}$Ni could be seen: The Solar Maximum Mission \citep{Simnett:1981} and its Gramma-Ray Spectrometer instrument \citep{Ryan:1979} showed the characteristic lines from decay of $^{56}$Co at 847 and 1238 keV, respectively \citep{Matz:1988}. SN2014J was the first such Type Ia supernova where those characteristic $\gamma$~rays have been seen; this measurement and its implications is discussed in detail in the last Section of the Chapter below. 

\begin{figure}
\centerline{
  \includegraphics[width=0.6\textwidth]{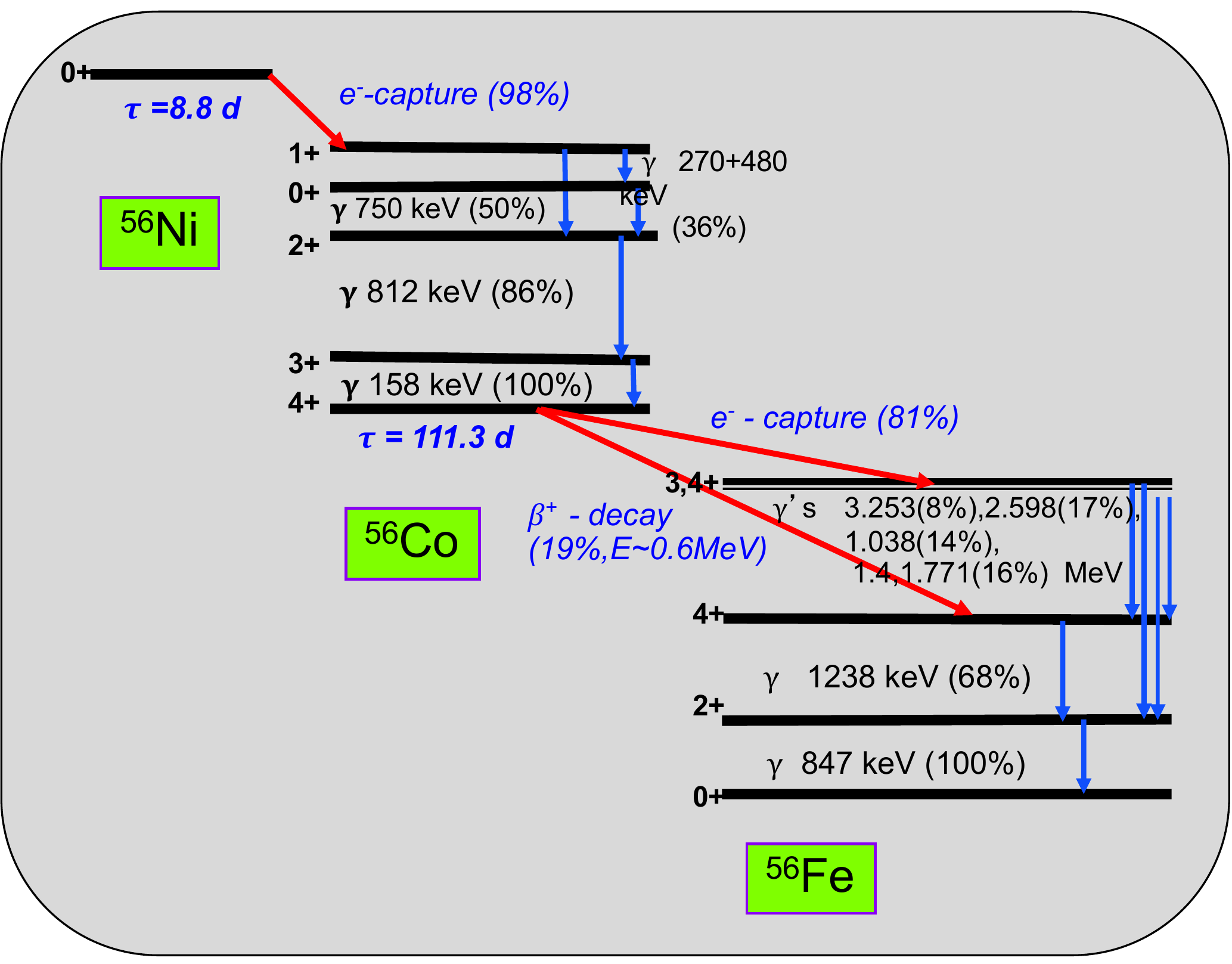}}
  \caption{\Ni decay, as an example of how the radioactive decay in supernova envelopes provides the power for subsequent thermal radiation, making supernovae being bright astronomical objects. Brightness is fading due to radioactive decay and dilution of the absorber for particles carrying the energy from radioactive decay.  }
  \label{fig:56Ni-decay}
\end{figure}

The energy released from radioactive decay into its surroundings provides astronomical opportunities,  observing high-energy photons (as covered in this Handbook) from energized material, where other sources of energy are absent or implausible. 
The prominent case here are supernova explosions: Under explosive conditions in both physical types of supernova explosions, nuclear matter is processed in near-nuclear reaction equilibrium. This equilibrium aims at a balance of number of particles (the \emph{phase space} factor, according to Liouville's theorem), and the minimisation of kinetic energy as binding energies per nucleon are maximised. 
Under most conditions, this favours nucleon binding in the form of the radioactive nucleus $^{56}$Ni, for matter composed of equal numbers of protons and neutrons (symmetric matter). $^{56}$Ni decays first within 8.8 days to radioactive $^{56}$Co, which again decays within 111 days to the end product, stable $^{56}$Fe; this is illustrated in Figure~\ref{fig:56Ni-decay}. The total energy released per $^{56}$Ni nucleus in this decay chain is 6.7~MeV; for 1~\Msol of $^{56}$Ni, this corresponds to an energy of 2.3~$\times$10$^{50}$erg. 
The final Section of the next Chapter 
below discusses specific astrophysical applications.

We note another prominent example where release of nuclear energy has important consequences:  the decay of radioactive $^{26}$Al embedded in planetesimals or dust grains of a protostellar nebula. Being formed out of cold interstellar matter at typical temperatures of order 10~K, these smallest solids contain all material that eventually may end up to form a planet. The decay of $^{26}$Al liberates a heating power of 0.5~mW~g$^{-1}$, which is sufficient under many conditions to heat a protostellar solid body to temperatures resulting in outgassing of volatile components, and in particular of water \citep{Lichtenberg:2019}.  This, however, is not within the scope of this Handbook.  

Finally, the energy in radioactive decay is mostly released in the form of characteristic $\gamma$~rays, from de-excitation of the daughter nucleus, in addition to kinetic energies of the daughter nucleus and leptons that result from the decay. These $\gamma$~rays are accessible to direct measurement, and provide an astronomy of radioactive cosmic materials. Examples are the measurements of characteristic $^{56}$Ni decay in supernova SN2014J, of $^{44}$Ti decay in the young supernova remnants of Cas A and SN1987A, and the diffuse $\gamma$-ray glow of our Galaxy from $^{26}$Al; all of these will be discussed in more detail below. 

\subsection{Astrophysical studies using radioactivity} 

\subsubsection*{Tracing past activity}  
\label{sec:pastActivity}
 
\begin{figure}  
\centering
\includegraphics[width=\columnwidth]{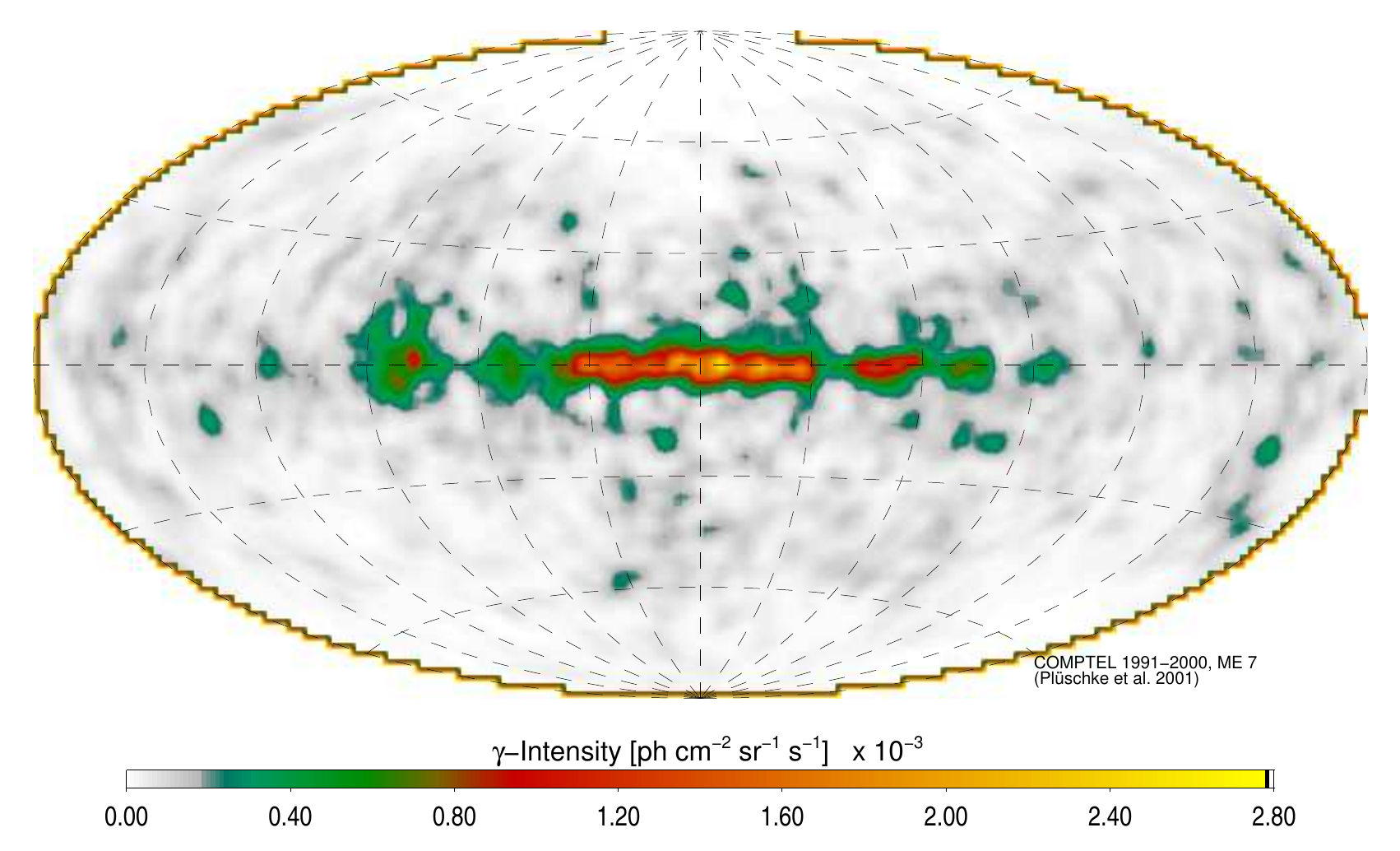}
\caption{The all-sky emission of $\gamma$~rays from radioactive decay of $^{26}$Al.  This image \citep{Pluschke:2001c} was obtained using a maximum-entropy regularization together with the maximum-likelihood method to iteratively fit a best image to the measured photon events.}
\label{fig_Al26map}
\end{figure}   

$^{26}$Al was the first unstable isotope which was detected to decay in the interstellar medium \citep{Mahoney:1982}. Its lifetime (1~My)  is shorter than typical times of stellar evolution (Gyrs), and much shorter than the age of the Galaxy itself or older stars herein ($>$10~Gy). So it must have been produced recently.
This observation of direct characteristic $\gamma$~rays from radioactive decay of the $^{26}$Al isotope demonstrated that nucleosynthesis, in some unknown mix of cosmic sources that include AGB stars \citep{Iben:1983} and WR stars \citep{Meynet:1994b}, novae \citep{Clayton:1974b}, core-collapse supernovae \citep{Timmes:1995}, and cosmic-ray reactions \citep{Kozlovsky:1987}, must have happened within the recent few Myrs of Galactic history. 

The diffuse emission from \Al decay (Figure~\ref{fig_Al26map}) was measured in detail during the first sky survey in $\gamma$~rays towards the turn of the century, with the COMPTEL telescope \citep{Schoenfelder:1993} on NASA's Compton $\gamma$-ray Observatory \citep{Gehrels:1993a}, in its 1991-2000 mission. 
This measurement made a meaningful imaging analysis possible for the first time \citep{Diehl:1995,Diehl:1995b,Knodlseder:1999}, and a sky map of the \Al emission with a resolution of about 4~degrees was derived through maximum-entropy analysis, shown in Figure~\ref{fig_Al26map} \citep{Pluschke:2001c}.
Apparently, \Al is of large-scale, galactic, origins. The theoretical considerations were put together with these observational results to conclude that massive stars and their core-collapse supernovae were believed to dominate the \Al production in the Galaxy \citep{Prantzos:1996a}.

A main supporting argument for the Galaxy-wide contributions seen in \Al radioactivity was contributed from ESA's INTEGRAL mission \citep{Winkler:2003} and its imaging $gamma$-ray spectrometer SPI \citep{Vedrenne:2003}. SPI detectors with a high spectral resolution of about 3~keV enabled the observation of the characteristic signature of large-scale Galactic rotation in spectroscopy of the $\gamma$-ray line from \Al decay \citep{Diehl:2006d,Kretschmer:2013}, that had been predicted already in 1978 \citep{Lingenfelter:1978}.
Therefore, the observed gamma-ray flux can be translated into an observed total emitting mass of \Al within our Galaxy. This makes use of geometrical models of how sources are distributed within the Galaxy, such as double-exponential disks and spiral-arm models \citep[see][for details]{Diehl:2006d}. 
A first mass estimate of 2.8$\pm$0.8~\Msol \citep{Diehl:2006d} was in line with theoretical expectations.
Having this \Al mass value, one can now employ theoretical models for the dominating massive stars and their supernovae and their \Al yield, and in this way obtain the core-collapse supernova rate that is required to produce this much $^{26}$Al. The value obtained in this way  \citep{Diehl:2006d} was a core-collapse supernova rate of 1.9$\pm$1.1 events per century for our Galaxy. The relatively-large uncertainty quoted herein accounts for uncertainties both in the geometrical model (impacting on the \Al mass estimate) and in the core-collapse supernova modeling. Nevertheless, in this way, one had, for the first time, a supernova rate estimate that relied on observations encompassing the entire Galaxy, rather than inferences from other galaxies or from star counts near the solar system \citep[see][for a discussion of these details and alternate methods]{Diehl:2006d}.
The supernova rate is key to driving turbulence within the interstellar medium \citep{Krumholz:2018,Koo:2020}, hence a key parameter to understand the dynamical state of interstellar medium.

The \Al mass estimate was revised and refined, as better geometrical models for the assumed source distribution in the Galaxy could be developed, also to account for foreground emission from more-nearby massive-star groups, which reduces the estimated mass of \Al in the Galaxy. 
The total mass of \Al in the Galaxy is now estimated to be between 1.8 and 2 ~\Msol\   \citep{Diehl:2018f,Pleintinger:2020}. 
The value of the Galactic core-collapse supernova rate therefore was updated with better estimates of the \Al mass and of model yields to 1.4$\pm$1.1~century$^{-1}$ \citep{Diehl:2018f,Pleintinger:2020}, or one such supernova in our Galaxy every 71 years.

At face value, the 294 supernova remnants observed in the Galaxy \citep{Green:2019} thus present a tension with this value \citep[see][for a discussion of the astrophysical issue]{Chomiuk:2011}, as they would suggest a maximum sampling age of 21000~y, while significantly larger ages have been inferred for some of these remnants. This, however, may be just another illustration of the dependence of the supernova remnant appearance on their surroundings.

With $^{44}$Ti, another radioactive isotope is attributed to supernovae, but it has a much shorter radioactive lifetime of just 86 years \citep{Ahmad:2006}. 
Upon decay to $^{44}$Sc, two lines are emitted at 69 and 78~keV as $^{44}$Sc decays to its ground state; $^{44}$Sc decays again within 5 hours to $^{44}$Ca, whose de-excitation emits a $\gamma$-ray line at 1156~keV energy.
$^{44}$Ti decay $\gamma$-rays have been observed from three supernovae, as discussed below in the Section on Diagnostics from explosions. But with this shorter lifetime, few sources across the Galaxy are expected to be found, even if one assumes that each supernova would eject radioactive $^{44}$Ti. The connections between the rate of supernovae in our Galaxy and the number of sources that could be found through $^{44}$Ti radioactivity have been analysed \citep{The:2006,Dufour:2013}, with the result that just a few $^{44}$Ti sources are expected.
The finding of the 360-year old Cas A supernova remnant as the only such source in the Galaxy hence is remarkable. Searches have been made with $\gamma$-ray telescopes such as both COMPTEL \citep{Dupraz:1997} and both INTEGRAL main telescopes IBIS \citep{Renaud:2006,Tsygankov:2016} and SPI \citep{Weinberger:2020}. No new sources have been found by either of these, with $\sim$ 1-2 debated marginal candidates, such as Vela Junior \citep[][]{Iyudin:1998,Schonfelder:2000b,Slane:2001,Weinberger:2020}.  
In conclusion, it is asserted that $^{44}$Ti ejection and $\gamma$-ray emission is attributed to a subclass of core-collapse supernovae only, and requires special circumstances of the supernova explosion (as discussed below in the last Section of the next Chapter).

Also  $^{60}$Fe $\gamma$-rays from its characteristic radioactive decay have been observed, from a cascade transition to the $^{60}$Co ground state with $\gamma$ rays at 1332 and 1173~keV. $^{60}$Fe  with $\tau=$3.6~Myrs is a second radio-isotope that is suited to trace accumulated past nucleosynthesis activity in the recent history of our Galaxy. 
A first signal from $^{60}$Fe decay had been found with the scintillation detectors of the Reuven Ramaty High-Energy Solar Spectroscopic Imager (RHESSI) mission \citep{Lin:2002} that was aimed at the Sun for solar flare observations. Being pointed at the Sun, observations of celestial $\gamma$~rays from the Galaxy, i.e. $^{26}$Al and $^{60}$Fe, were serendipitous, as the sky passes through the field of view of the unshielded detectors as a background varying over time \citep{Smith:2003}. Equipped with high-resolution Ge detectors, a clear signal from diffuse Galactic $\gamma$-ray emission from $^{26}$Al and $^{60}$Fe was found \citep{Smith:2003}. 
But spectral resolution was lost over the first few years of the 2002 -- 2018 mission, so that it became more difficult to recognise the corresponding lines above background. 
The SPI spectrometer on INTEGRAL also collected data since 2002, and also found the signal from $^{60}$Fe decay \citep{Harris:2005,Wang:2007}. 
Unlike for RHESSI, SPI detectors were periodically heated to achieve an \emph{annealing} of the degradation of spectral resolution from cosmic-ray bombardment in space. But it had been difficult for both instruments, with significances not exceeding the 5$\sigma$ level, as spacecraft activation of $^{60}$Co from cosmic rays occurs with time, and as the $^{60}$Fe $\gamma$-ray brightness is much below that of $^{26}$Al. The latter is quite in contrast to theoretical predictions \citep[e.g.][]{Timmes:1995,Rauscher:2002,Woosley:2007b}. 

The standard hypothesis is that both $^{26}$Al and $^{60}$Fe synthesis are dominated by massive stars, where $^{60}$Fe is synthesised in the burning He and C shells from neutron capture on pre-existing Fe nuclei, and released with the supernova explosion into the interstellar surroundings \citep{Chieffi:2002,Limongi:2006c,Woosley:2007b,Limongi:2018}. 
But rare types of supernovae could be significant, too \citep{Woosley:1997}. 
It was important, therefore, to exploit the $^{60}$Fe $\gamma$-ray signal and discriminate potential single-source origins from diffuse emission, which was obtained by comparing sets of sky models fitting INTEGRAL/SPI data \citep{Wang:2020}. This confirmed the diffuse nature of the $^{60}$Fe $\gamma$~rays, and also provided a best constraint on  $^{60}$Fe $\gamma$-ray intensity of below 0.4 of the  $^{26}$Al $\gamma$-ray brightness, considering also systematic uncertainties from model fitting and backgrounds at best \citep{Wang:2020}.

This ratio of two radioactivities both originating from massive stars is important, as it constrains the interior processes in such massive stars, independent of their number and locations as these cancel out in the ratio (see discussion in \citep{Woosley:2007b}, for example). Understanding $^{60}$Fe ejection from typical core-collapse supernovae then provides the background knowledge to interpret the $^{60}$Fe nuclei that have been found in sediments on Earth, and also in lunar probes and in Antarctic snow \citep{Wallner:2021}: 
These important findings of radioactive material deposits are proof of specific supernova activity near our solar system in the past few Myrs \citep{Ellis:1996}. 
Exciting quantitative interpretations of this are the dating of such events and their relations to other known objects and their presumed historic evolution at distances of order 10 to 100 pc only \citep{Wallner:2016,Breitschwerdt:2016}. But these rely on the combination of theoretical modelings of the sources and their generic, Galaxy-wide, confirmation, as provided by the diffuse $^{60}$Fe $\gamma$-ray lines . 

Note that radioactivities are important tools for cosmic age-dating and for diagnostics of previous exposure of materials to high-energy reactions, in the fields of meteorites, of stardust, and of cosmic rays. These are not discussed in this Handbook; see  \citep[]{Clayton:1988b,Clayton:2004,Zinner:2008,Israel:2018} for reviews.

\subsubsection*{Tracing flows of nucleosynthesis ejecta}  
\label{sec:flows}

\begin{figure}  
\centering
\includegraphics[width=0.6\columnwidth]{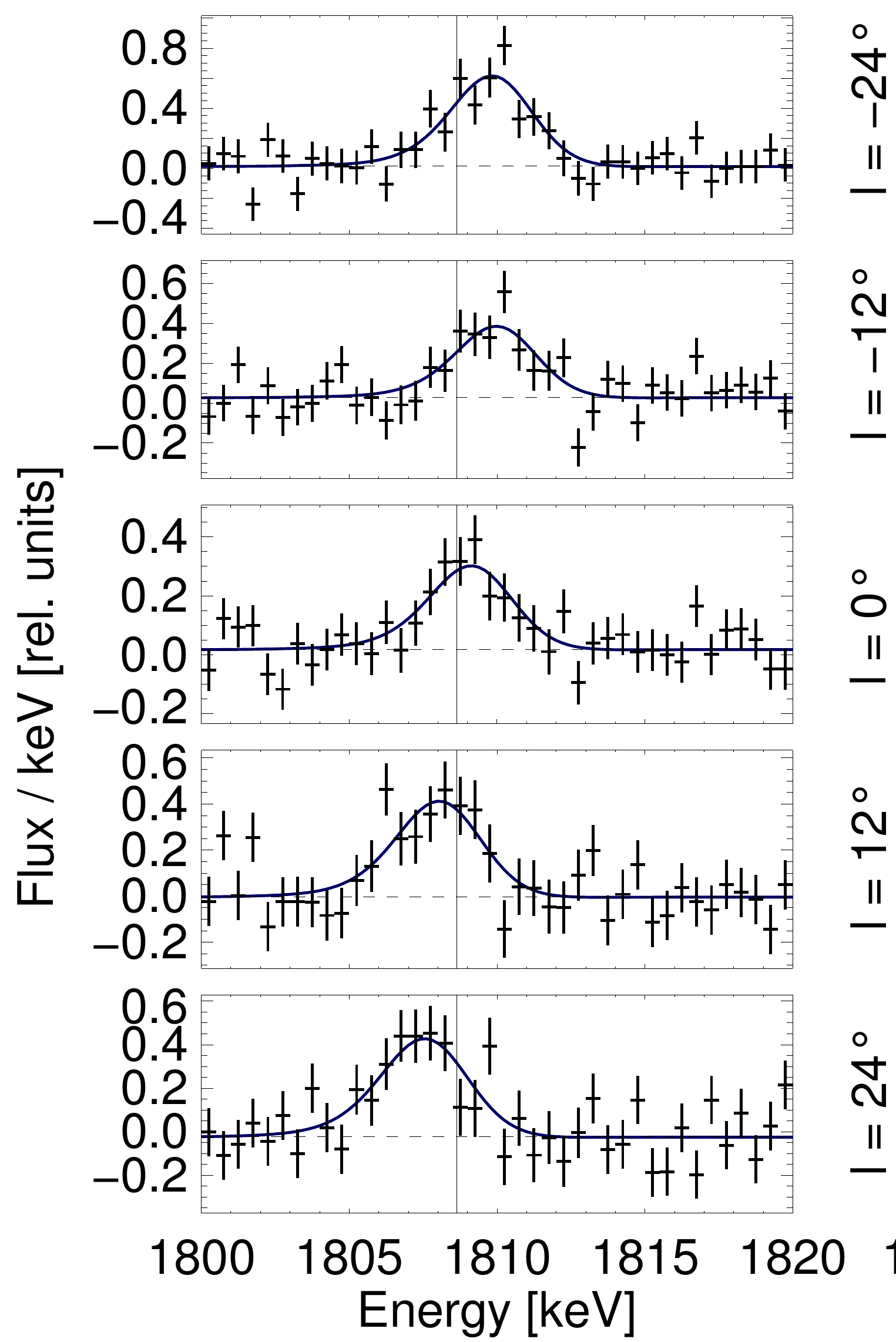}
\caption{The \Al line as seen towards different directions (in Galactic longitude). This demonstrates kinematic line shifts from the Doppler effect, due to large-scale Galactic rotation \citep{Kretschmer:2013}. }
\label{fig_Al26_longitudes}
\end{figure}   
 
Some radioactive isotopes have a lifetime approaching the typical recycling time scale of interstellar gas into stars of 10$^7$ to 10$^8$ years. This offers a way to \emph{trace} the flow of nucleosynthesis ejecta directly, i.e. through their radioactive afterglow. Note that other astronomical signatures of the release of newly-produced cosmic material are of rather short duration, by comparison; supernova remnants, e.g., remain astronomically visible for times of order of several 10,000~years only \citep{Koo:2020,Vink:2012,Reynolds:2008}. 
\Al with its lifetime of 1.04$\times$10$^6$~y is on the short side of the recycling time of cosmic gas, but offers the brightest emission for this purpose.
Its detection, measurements, and global Galactic interpretations in terms of massive-star feedback have been discussed in the previous Section.

In 1995, the GRIS balloon experiment \citep{Tueller:1988} had reported an indication that the \Al~line was significantly broadened. A value of  6.4~keV \citep{Naya:1996} was obtained from a measurement with the high-resolution Ge detectors employed by this instrument. 
If interpreted as kinematic Doppler shifts of astrophysical origin, this translates into a \Al motion of 540~km~s$^{-1}$ \citep{Naya:1996}. 
Considering the 1.04$\times$10$^6$~y decay time of $^{26}$Al, such a large velocity observed for averaged interstellar decay of \Al would naively translate into kpc-sized cavities around \Al~sources, so that velocities at the time of ejection would be maintained during the radioactive lifetime. An alternative hypothesis is that major fractions of  \Al condensed onto grains, which would maintain ballistic trajectories in the tenuous interstellar medium \citep{Chen:1997,Sturner:1999}. 

\begin{figure}[t]  
\centering 
\includegraphics[width=0.8\columnwidth]{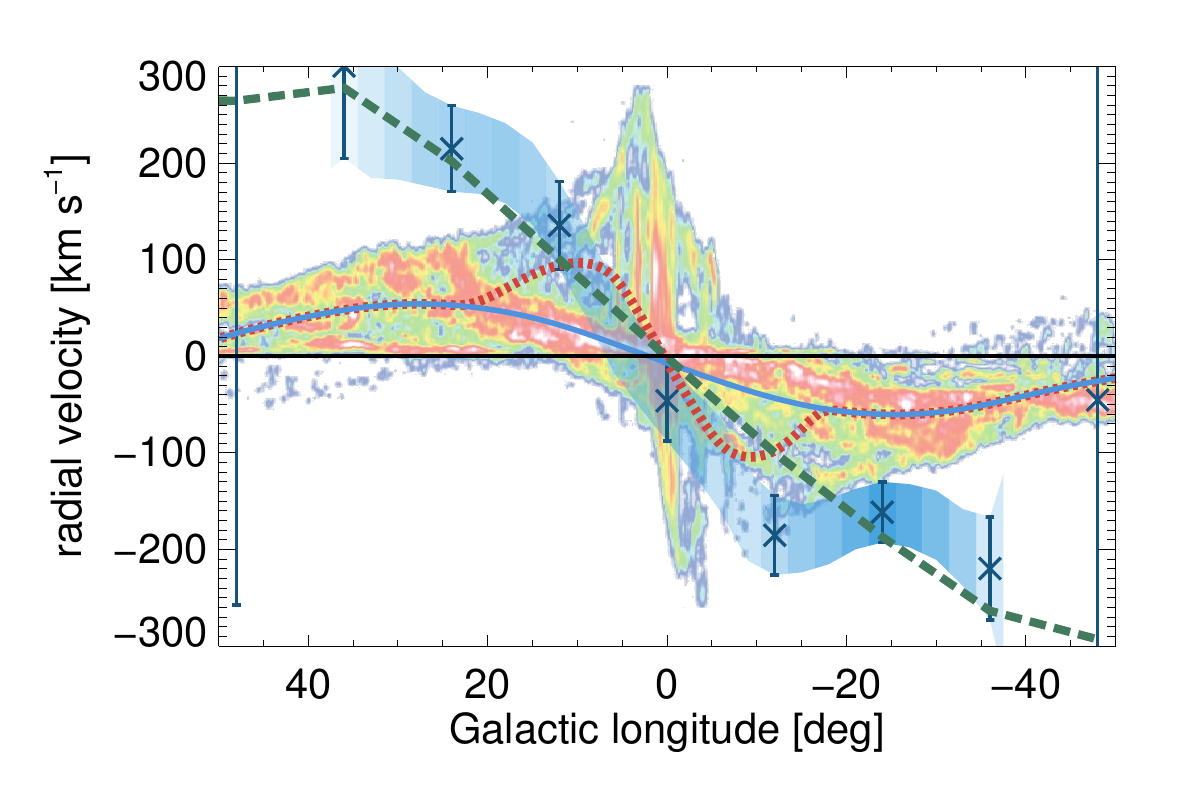}
\caption{The line-of-sight velocity shifts seen in the \Al\ line versus Galactic longitude (data points with error bars), compared to measurements for molecular gas in CO (colors, intensity-coded from blue to red). The dashed line represents a model from \Al decay into cavities at the leading edge of spiral arms, as shown in Figure~\ref{fig:spiralarmbubbles}.  \citep{Kretschmer:2013}. }
\label{fig:al_long-velocity} 
\end{figure}

More recent high-quality spectroscopic data from INTEGRAL's $\gamma$-ray spectrometer SPI have deepened and detailed these observations. SPI maintains a resolution of 3~keV at the energy of the \Al line (1809~keV) for multi-year data accumulation \citep{Diehl:2018}; this corresponds to a Doppler velocity shift resolution of about 100 km~s$^{-1}$ for bright source regions \citep{Kretschmer:2013}. Additionally, the INTEGRAL spectrometer SPI also is an imaging instrument, thus capable of mapping the spectral properties of \Al emission across the Galaxy. 
Already predicted in 1978 as a signature of large-scale
 Galactic rotation \citep{Lingenfelter:1978}, this signal was then seen in Galactic-plane survey data from INTEGRAL/SPI. In early INTEGRAL results \citep{Diehl:2006d}, it appears as a blue shift when viewing towards the fourth quadrant (objects on Galactic orbits approaching) and a red shift when viewing towards the first quadrant (receding objects, on average).
The consolidated signature, with more years of exposure, is shown in Figs.~\ref{fig_Al26_longitudes} and \ref{fig:al_long-velocity}).

\begin{figure}[t]  
\centering 
  \includegraphics[width=0.5\linewidth]{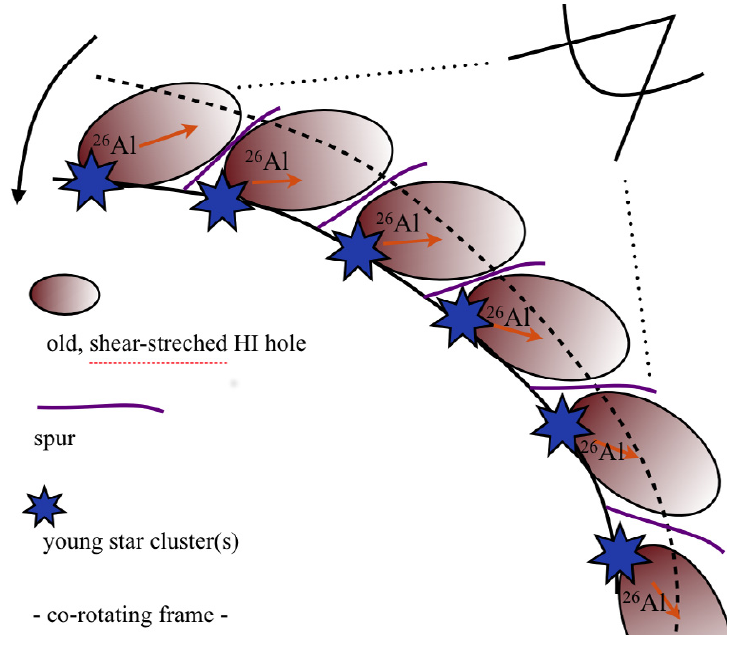}
\caption{A model for the different longitude-velocity signature of $^{26}$Al, assuming \Al\ blown into inter-arm cavities at the leading side of spiral arms \citep{Krause:2015}. }
\label{fig:spiralarmbubbles} 
\end{figure}   

Comparison of the observed \Al velocity from large-scale Galactic rotation \citep{Kretschmer:2013} with the velocity of molecular gas exhibits a puzzling discrepancy (Fig.~\ref{fig:al_long-velocity}).
The velocities seen for \Al throughout the plane of the Galaxy   \citep{Diehl:2006d,Kretschmer:2013} exceed the velocities measured for  molecular clouds, stars, and (the most precisely measured velocities of) maser sources by typically as much as 200~km~s$^{-1}$ (see Fig.~\ref{fig:al_long-velocity}). 
This high apparent bulk motion of decaying $^{26}$Al means that the velocities of these radioactive nuclei remain higher than the velocities within typical interstellar gas for 10$^6$ years. Additionally, Figure~\ref{fig:al_long-velocity} shows that there is an apparent bias for this excess average velocity in the direction of Galactic rotation. 

This has been interpreted \citep{Krause:2015,Krause:2021} as $^{26}$Al decay occurring preferentially within large cavities (superbubbles), which are elongated into the direction of large-scale Galactic rotation (Fig.~\ref{fig:spiralarmbubbles}). 
If such cavities are interpreted as resulting from the early onset of stellar winds in massive-star groups, they characterise the source surroundings at times when stellar evolution terminates in  core-collapse supernovae. Such  wind-blown superbubbles around massive-star groups plausibly extend further in space in forward directions and away from spiral arms (that host the sources), as has been seen in images of interstellar gas from other galaxies  \citep[][and references therein]{Schinnerer:2019}.  
Such superbubbles can extend up to kpc \citep{Krause:2015} \citep[see also][]{Rodgers-Lee:2019,Krause:2021,Nath:2020}), which allows \Al to propagate at velocities similar to the sound speed within superbubbles. 
The dynamics of such superbubbles into the Galactic halo above the disk are unclear; this is perpendicular to the line of sight, so that measurements of \Al line Doppler velocities cannot provide an answer \citep[see][for a discussion of outflows into the halo]{Krause:2021}.
Nevertheless, these measurements underline the important consequences of massive-star clustering in shaping the interstellar medium, with connections to the astrophysics of stellar feedback in general \citep[see][for recent theoretical considerations]{Krause:2020,Chevance:2022}.

One particular such superbubble has been identified near the solar system and towards the Orion region: 
The Eridanus cavity has been recognised in diffuse X-ray emission \citep{Burrows:1993}, its boundaries are delineated in HI radio emission \citep{Heiles:1999}, and $\gamma$~rays from \Al have been detected \citep{Siegert:2017a} 
\citep[Implications of these observations in terms of massive-star feedback have been discussed, a.o., by][]{Fierlinger:2016}.

Radioactive decay is often accompanied by the emission of positrons, if the decay is a $\beta^+$~decay; \Al decay as shown in Figure~\ref{fig:26Al-decay} is an example. 
Once having escaped from the source, positrons will propagate through the interstellar medium as directed by magnetic fields, limited in time by interactions with particles and fields along the way. One of these interactions, in this case, is the annihilation of the positron with its anti-particle, the electron. This may occur in different ways, either directly, or through radiative captures, or through charge-exchange collisions with hydrogen atoms forming an intermediate positronium atom. As a result of positron annihilation, characteristic $\gamma$~rays are generated, with a pair of photons at 511~keV energy for direct annihilations and for annihilation through the intermediate formation of para-positronium, and a spectrum rising in energy with an upper energy limit of 511~keV if through the intermediate formation of ortho-positronium, as spins of positronium and the annihilation $\gamma$-rays must be balanced.
The study of positron propagation from their nucleosynthesis sources through interstellar surroundings  \citep{Alexis:2014} suggests that propagation out to at most a few 100~pc may occur. 
The $\gamma$-ray emission from positron annihilation has been measured in detail and imaged across the entire sky with the INTEGRAL mission and data from the SPI instrument \citep{Knodlseder:2005,Jean:2006,Siegert:2016,Churazov:2020}. 
It is found that the emission morphology of positron annihilation $\gamma$~rays is very different from those of any of the radioactivity candidates to produce the positrons, even if propagation is accounted for \citep[see, e.g.][]{Martin:2010,Prantzos:2011}. 
Therefore it is concluded that radioactivity only contributes a minority of positrons that are seen annihilating throughout the Galaxy; see \citep{Prantzos:2011} for a review of positron astrophysics with lessons and remaining puzzles.

\subsubsection*{Diagnostics of explosions}  
\label{sec_explosions}

Explosive nucleosynthesis occurs in complex networks of nuclear reactions. As the explosion is launched and proceeds, the environment and conditions for nuclear reactions vary due to the complex dynamics of material, and these are reflected in the products that result from a supernova explosion \citep{Arnett:1996a}. Therefore, measurements of the radioactive ejecta of supernova explosions can be used directly as a diagnostics of the inner explosion, complementing sparsely-available measurements from neutrinos and gravitational waves, which are the only messengers that directly escape from these inner regions of a supernova explosion. By comparison, all low-energy radiation from X~rays down to infrared and radio are indirect, and shaped by processes outside of these inner regions.
 
The gamma-rays and positrons emitted from radioactive decay chain of $^{56}$Ni through $^{56}$Co (decay time $\tau$=8.8 days)  to $^{56}$Fe ($\tau$=111~days) energise the ejecta and envelope of the supernova from inside. The scattered and re-processed radioactive energy is then responsible for the light that appears at the \emph{photosphere} and makes supernovae to be bright sources. 
This photospheric emission has been key to many astrophysical studies of supernova explosions, in spite of its indirect origins far outside the explosion physics, because observations have been readily made available by the ensemble of telescopes world-wide and from infrared to UV wavelengths. 
For example, empirical laws have been used to estimate the amount of $^{56}$Ni produced in the explosion \citep{Arnett:1982}, and spectroscopic measurements of elemental abundances have been used to infer the neutron to proton ratio in the explosion \citep{Brachwitz:2000,Mori:2018,Yamaguchi:2014}.  
But if accessible, the radioactivity also  provides a more-direct diagnostic of the explosions themselves.

The direct measurement of $\gamma$~rays of $^{56}$Ni decay had been predicted from supernova models \citep[e.g.][]{Colgate:1969, Clayton:1974}.
But it took 40 years until, for the first time, the radioactive energy injection of $^{56}$Ni decay could be measured directly as it powers the light from thermonuclear supernovae.  INTEGRAL's instruments in space orbit made this possible \citep{Churazov:2014}, as SN2014J occurred on January 22, 2014 \citep{Fossey:2014aa} at a distance of only 3.3~Mpc in the nearby starburst galaxy M82 \citep{Foley:2014}. 
INTEGRAL's spectrometer SPI contributed $\gamma$-ray spectroscopy at fine resolution of the $^{56}$Ni decay lines (see Figure~\ref{fig:56Ni-decay}) \citep{Diehl:2014,Diehl:2015,Isern:2016}. This is a valuable complement, as it bypasses the uncertainties and complexities of radiation transport within the supernova, that downscatters MeV radiation by many orders of magnitude into optical light. Its diagnostic power had been emphasised by modeling work \citep[e.g.][]{Summa:2013} long before SN2014J occurred.

The amount of $^{56}$Ni inferred from the $\gamma$-ray flux of 0.49$\pm$0.09~\Msol\  \citep{Diehl:2015} is in agreement with the amount inferred from the optical brightness of the supernova, as based on the empirical peak-brightness/$^{56}$Ni heating-rate relation discussed above (\emph{Arnett's rule} \citep{Arnett:1982}). 
The $^{56}$Ni mass determination with infrared light curves \citep{Dhawan:2016} also is in agreement with this direct $\gamma$-ray based $^{56}$Ni mass determination in SN2014J.
A key ingredient of modeling supernova light from thermonuclear supernovae in the broader electromagnetic spectrum from $^{56}$Co radioactivity with its characteristic 111-day decay time, has thus been measured by most-direct messengers,  
as a reassuring confirmation of our standard understanding of the origins of supernova light. 

\begin{figure}
  \centering
  \includegraphics[width=0.8\columnwidth]{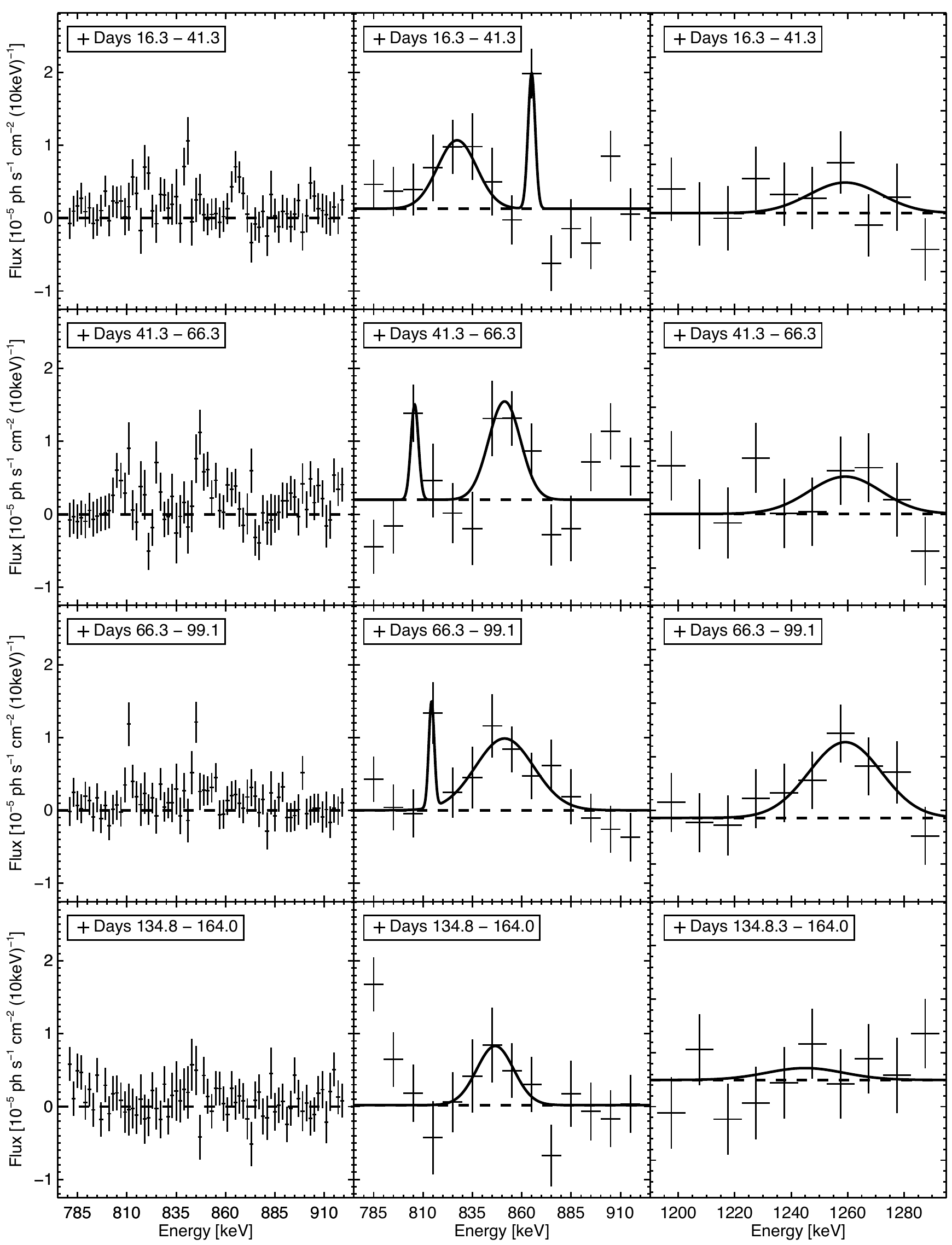}
   \caption{SN2014J signal intensity variations for the 847 keV line ({\it center}) and  the 1238 keV line ({\it right}) as seen in four epochs of high-resolution $\gamma$-ray observations, in 10 keV energy bins.  Clear and significant emission is seen in the lower energy band ({\it left and center}) through a dominating broad line attributed to 847~keV emission, the emission in the high-energy band in the 1238~keV line appears consistent and weaker, as expected from the branching ratio of 0.68  ({\it right}). 
   For the 847 keV line, in addition a high-spectral resolution analysis is shown at 2 keV energy bin width ({\it left}), confirming an irregular appearance, i.e. not homogeneously in the form of a broad Gaussian.  \citep[From][]{Diehl:2015}.}
  \label{fig_SN2014J_SPI-spectra-set-847-1238}
\end{figure}

In details of the $\gamma$-ray spectroscopy, there had been indications of interesting irregularities, that may be washed out by the processes of radiation transfer in a supernova envelope.
The $\gamma$-ray line emission from decay of $^{56}$Co was traced over 3 months with the INTEGRAL $\gamma$-ray spectrometer.
Thus the line centroid and width could be constrained in their evolution, as shown in Figure~\ref{fig_SN2014J_SPI-spectra-set-847-1238}. 
Naively, one would have expected a Doppler-broadened Gaussian line to appear and fade in its brightness, with some centroid shift from blue to red as the facing ejecta would shine early and receding ejecta from the distant part of the supernova would add later.
But, as shown in Figure~\ref{fig_SN2014J_SPI-spectra-set-847-1238}, the spectra rather show surprising spikes, which appear to come and go with time. Although statistical noise is a concern, it was asserted that a smooth appearance of the lines as suggested from a spherically-symmetric gradual transparency to centrally located $^{56}$Ni could be excluded \citep{Diehl:2015}. 
Rather, it is indicated that individual clumps of $^{56}$Co decay at different bulk velocities may have appeared at different times, signifying substantial deviations from spherical symmetry of explosion or $^{56}$Co distribution  \citep[see][for more detail on the $^{56}$Co $\gamma$-ray signal]{Diehl:2015}.
Thus, $\gamma$-ray spectroscopy provides an additional indication that non-sphericity may be significant in type Ia explosions, and rather smoothed out in signals based on bolometric light re-radiation from the entire supernova envelope.   

\begin{figure}
  \centering
  \includegraphics[width=0.6\columnwidth]{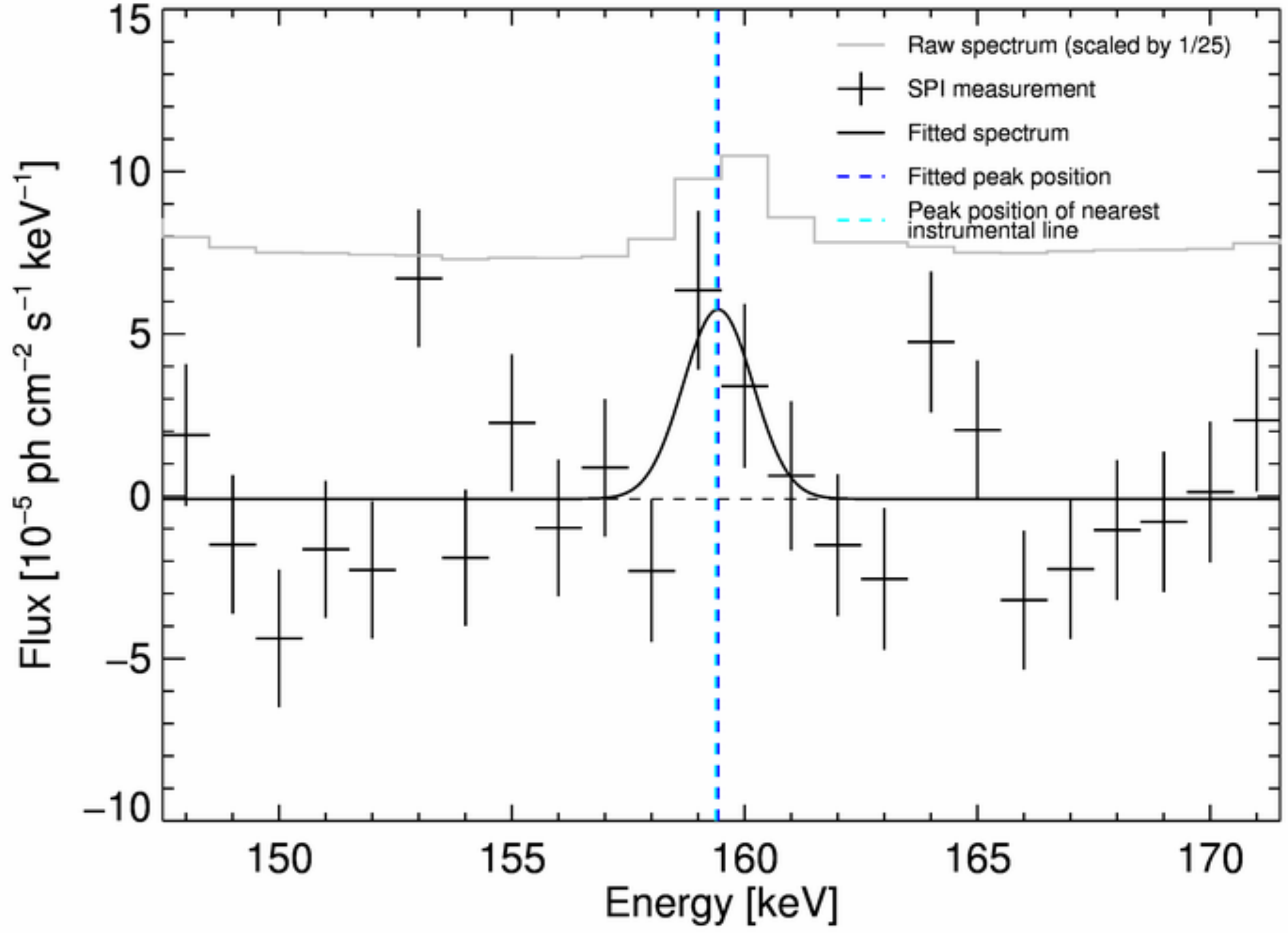}
   \caption{Early $\gamma$-ray spectrum from SN2014J, finding the characteristic line at 158 keV from \Ni decay. Observed from a three-day interval around day 17.5 after the explosion, this confirms an early visibility of \Ni, probably close to the surface, rather than embedded in the supernova center. The SPI instrumental background is shown as a scaled histogram, showing the SN2014J line offset from the centroid of a strong background line. The measured intensity corresponds to an initially-synthesised \Ni mass of 0.06 \Msol. \citep[From][]{Diehl:2015}}
  \label{fig_SN2014J_SPI-spectrum_158}
\end{figure}

In SN2014J, even more spectacular was the detection of $^{56}$Ni decay lines early-on (Figure~\ref{fig_SN2014J_SPI-spectrum_158}): It was believed that $^{56}$Ni would always be embedded so deeply within the supernova's core that even $\gamma$ rays could not leak out before $^{56}$Ni was converted to $^{56}$Co, due to its radioactive lifetime of about 9 days. Only some \emph{He cap} models included the possibility of early $\gamma$-ray emission from $^{56}$Ni decay \citep{The:2014}, as helium deposition on the surface of the white dwarf could cause a helium surface explosion triggering the thermonuclear supernova. Thus, this discovery of early $^{56}$Ni $\gamma$-ray lines was discussed as a support for such a double detonation \citep[see][for more detail]{Diehl:2014}.

We now turn to the case for core-collapse supernovae. 
Observations of supernova SN1987A with $\gamma$-ray telescopes had shown that the radioactive power source of supernova light was embedded within the massive envelope of such a supernova somewhat different than the standard \emph{onion shell} picture of a massive star at the end of its stellar evolution would suggest: 
The SMM Gramma-Ray Spectrometer  showed a surprisingly-early appearance of  the characteristic lines from decay of $^{56}$Co at 847 and 1238 keV \citep{Matz:1988,Leising:1990}, only six months after the explosion. 
Detailed line spectroscopy of this signal suggests an early redshift of the line centroids \citep{Leising:1990}, quite in contrast to expectations:  early inner radioactivity should be revealed from the near side of the supernova first. Followup spectroscopy with semiconductor detectors of higher spectral resolution than SMM's scintillation detectors then showed an expected evolution from early blue-shifted lines to  red shift of this early $^{56}$Co line signal \citep{Tueller:1990}. A detailed analysis of model ingredients that would be compatible with all $\gamma$-ray data on SN1987A $^{56}$Co line emission suggests that a rather particular combination of bulk initial velocity of the $^{56}$Ni produced, and some degree of non-spherical distribution with a suitable combination of envelope mass and explosion energy are needed  \citep{Jerkstrand:2020}. 

The bolometric light curve of SN1987A follows the predictions of fully trapping the energy from $^{56}$Co decay for several hundred days. It can be modulated by standard envelope density models and changes in the escape fractions of $\gamma$-rays due to decreasing densities as the envelope expands. This radiation escape only becomes a significant effect beyond about 1000 days after the explosion. 
At later times, the change of slope in the bolometric light curve indicates that power now is delivered from radioactivities with longer decay times, first $^{57}$Co and then $^{44}$Ti \citep{Seitenzahl:2014}. 
The $^{44}$Ti radioactivity as a power source of SN1987A emission at this time was then beautifully confirmed through observations of the characteristic decay lines directly.
$^{44}$Ti decays to $^{44}$Sc within $\tau$ \about~86~years \citep{Ahmad:2006}, emitting characteristic $\gamma$-rays of 68.87 and 78.36 keV from de-excitation of $^{44}$Sc. The subsequent decay of $^{44}$Sc to $^{44}$Ca occurs after $\tau$~\about~5.73~hours only, producing a characteristic $\gamma$-ray line at 1157.02 keV from de-excitation of $^{44}$Ca. Thus, for our purpose, we may characterise the decay of $^{44}$Ti with a decay time of 86~y and three characteristic lines at 69, 78, and 1157~keV energy. 
The 67.9 keV and 78.4 keV lines of $^{44}$Ti decay have been observed 25 years after explosion  with INTEGRAL instruments \citep{Grebenev:2012}, and more clearly \citep{Boggs:2015} with the NuSTAR hard X-ray telescope \citep{Harrison:2013}. 
The NuSTAR hard X-ray telescope provides a unique opportunity for observations of characteristic lines from radioactivity, as only up to energies of $\sim$80 keV it is possible to deflect and focus celestial photons with X-ray optics, thus increasing the collection area of a telescope beyond the detector surface area; this is common for standard telescopes, yet impossible at energies above 100 keV due to the penetrating nature of $\gamma$~rays.
Therefore, even at 55~kpc distance, the brightness of SN1987A was sufficient for a significant measurement of the $^{44}$Ti radioactivity  \citep{Boggs:2015}. The measured line flux of 3.5$\times$10$^{-6}$~ph~cm$^{-2}$s$^{-1}$ translates into a $^{44}$Ti amount of  1.5$\times$10$^{-4}$~\Msol.  
By comparison, the INTEGRAL result is somewhat debated, in particular with a strikingly-large inferred $^{44}$Ti amount of (3.1$\pm$0.8)$\times$10$^{-4}$~\Msol \citep{Grebenev:2012}; subsequent analyses of data with INTEGRAL's SPI instrument could not confirm a signal from SN1987A in these characteristic lines \citep{Weinberger:2020}. 
The characteristic lines from $^{44}$Ti decay in SN1987A are thus not bright enough for detailed spectroscopy. Therefore, a diagnostics is only provided by comparing measured fluxes to the yields of models of different types \citep[e.g.][]{Woosley:1991,Timmes:1996a,Nagataki:1998,Magkotsios:2008,Magkotsios:2010,Sukhbold:2016,Wongwathanarat:2017,Curtis:2019}. 
Note that these direct measurements all find $^{44}$Ti amounts that fall on the high side of theoretical expectations, which are generally in the range of a few 10$^{-5}$~\Msol  .

The second case of a core-collapse supernova with interesting observational detail in high-energy emission for supernova explosion diagnostics is the rather young supernova remnant Cas A. At a distance of 3.4~kpc and an age of 360 years \citep{Fesen:2006}, this remnant is close enough so that  telescopes at X- and $\gamma$-ray energies can measure all characteristic lines from decay of radioactive $^{44}$Ti at sufficient precision for supernova diagnostics.

\begin{figure}
\centering 
\includegraphics[width=0.7\columnwidth]{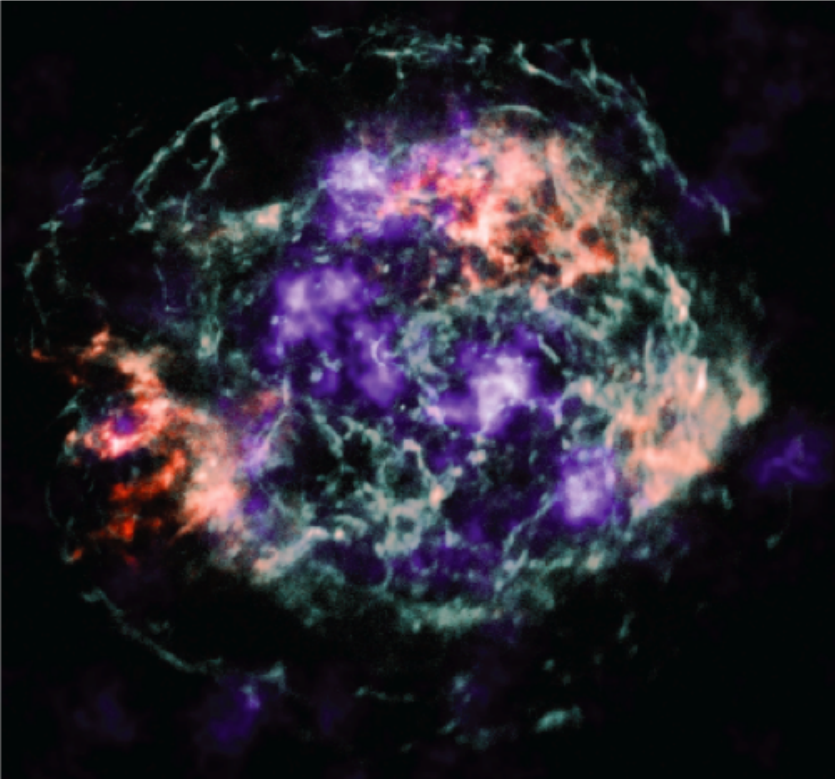}
\caption{The image \citep{Grefenstette:2014} of the Cas A supernova remnant demonstrates how radioactivity complements our view: Characteristic lines from $^{44}$Ti decay (blue) reveal the location of inner ejecta, while X-ray line emissions from iron (red) and silicon (green) atoms, that also are emitted from those inner ejecta, show a somewhat different brightness distribution, due to ionization emphasising the parts of ejecta that have been shocked within the remnant.}
\label{fig:CasAimage}
\end{figure}

The Cas A supernova remnant was the first source where $^{44}$Ti decay was directly observed through $\gamma$-rays with the COMPTEL instrument, and the characteristic line at 1157~keV \citep{Iyudin:1994}. Later this signal was confirmed by several other high-energy astronomy instruments, i.e. OSSE, RXTE,  \citep{The:1996,Rothschild:1999,Vink:2000,Siegert:2015}. 

With its size, it was ideally suited for the NuSTAR imaging hard X-ray telescope \citep{Harrison:2013}, to measure and image its radioactivity $\gamma$-rays from the radioactive $^{44}$Ti that had been ejected with the supernova 360 years ago \citep{Grefenstette:2014}. 
The image shown in Figure~\ref{fig:CasAimage} is an overlay of emission mapped with the Chandra X-ray telescope in characteristic lines from Fe and Si recombination lines, and emission from radioactive $^{44}$Ti decay in the 69 and 78~keV lines imaged with the NuSTAR multilayer mirrors. 
The recombination line image had been puzzling for a while, as it shows Fe emission located outside of Si structures. This should not be, if a massive star's iron core launches a core-collapse supernova with lighter elements further out. But the $^{44}$Ti radioactivity should be co-produced with any iron that may be ejected from the inner regions of the supernova. The NuSTAR data clearly show its location, which is near the center of the explosion, as expected.
The puzzle is resolved from the nature of the different line emissions: The recombination lines recorded with Chandra result as electron recombination occurs in highly-ionised plasma, and the atomic shell reaches its ground state, emitting characteristic lines in X~rays for these highly-ionised states. In contrast, radioactive decay is independent of ionisation state and occurs as nuclei are present, independent of density or temperature. Therefore, the iron recombination line emission is sensitive to biases from ionisation, and are even absent for fully-ionised plasma. Hence, the Chandra measurements shows iron recombining where it has been overrun by the reverse shock already, while iron in the interior of the remnant has not been ionised yet by the inward-moving reverse shock.

\begin{figure}
\centering 
\includegraphics[width=0.8\columnwidth]{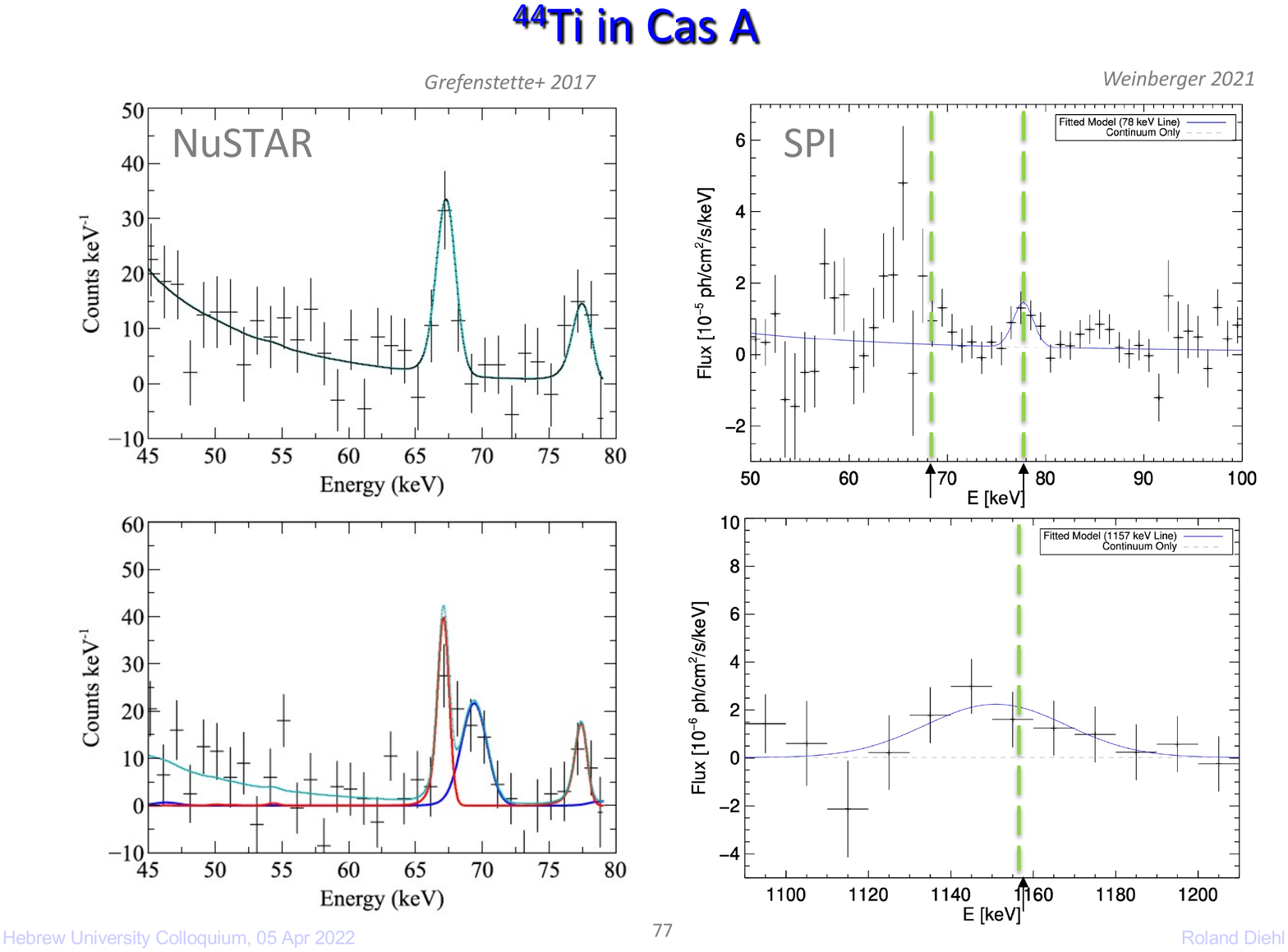}
\caption{The spectra for $^{44}$Ti emission from the Cas A supernova remnant show significant redshifts, indicating bulk motion away from the observer. The NuSTAR spectra for two different emission regions (\emph{left})  \citep{Grefenstette:2017} show that different regions show different bulk velocities, preferentially however away from the observer. The INTEGRAL/SPI spectra for all three lines (\emph{right}) \citep{Weinberger:2021} for emission intenrated across the entire remnant confirm this trend, and show in particular a clear Doppler shift for the 1157~keV line (\emph{lower right}).}
\label{fig:CasAspectra}
\end{figure}

Moreover, the image in $^{44}$Ti radioactivity lines
 spectacularly shows directly that radioactive ejecta appeared in several clumps, rather than as spherically-symmetric shells. 
After SN1987A's asphericity indicators (see above), this is another direct demonstration that sphericity is not common in core-collapse supernova explosions.
The NuSTAR measurement allowed a decomposition of the $^{44}$Ti signal through imaging spectroscopy, determine the spectra across the remnant for $\sim$~20 independent positions \citep{Grefenstette:2017}. 
These results show that there is kinematic diversity across the inner remnant and these nucleosynthesis ejecta, with remarkable redshift for a majority of the regions that could be discriminated (Figure~\ref{fig:CasAspectra}, left).
A deeper analysis of cumulative data of INTEGRAL/SPI consistently finds that all $^{44}$Ti decay lines are red-shifted by an amount that corresponds to a bulk velocity of (1800$\pm$800)~km~s$^{-1}$  (Figure~\ref{fig:CasAspectra}, right), consistent with the NuSTAR findings; SPI cannot resolve regions within the remnant and only provide an integrated spectrum. Note that the Doppler shift scales with energy; therefore, the data from the 1157~keV line contribute most-significantly to this bulk velocity determination with INTEGRAL.
Again, as for SN1987A, a remarkable deviation from sphericity is found for this core-collapse supernova that created the Cas A remnant.


Also fluorescent X-ray line emission at keV energies may be emitted from radioactive material, if decay occurs through electron capture and leads to such X-ray emission as the atomic-shell vacancy is replenished \citep{Seitenzahl:2015}; this may have added to the energisation of the late SN1987A light curve, but direct detections remain ambiguous \citep{Borkowski:2010}.

Thus, observations of radioactive decay $\gamma$-ray lines provide an important complement to the rich archives of supernova light curves and spectra at other wavelengths; each of these is at an opposite end of the complex radiation transfer within a supernova, with $\gamma$-rays from primary radioactive decay within the supernova and detailed spectra with elemental lines from emission of optical/IR at the supernova photosphere. 

\section{Summary}
Radioactive decay occurs as unstable nuclei are created in nuclear reactions of cosmic nucleosynthesis. The decay is mediated by the weak interaction, and hence largely independent of thermodynamic parameters such as density or temperature. Therefore, radioactive decay is an interesting emission process for characteristic $\gamma$~rays, which can provide new astronomical messages from sources of high-energy astrophysics. Mainly, the decay of radioactive isotopes signifies the past occurrence of such nuclear interactions to a sample of cosmic matter, also providing a clock which allows tracing back such history. The radioactive decay also provides a source of energy, which may be deposited to matter after typically a delay corresponding to the characteristic time of radioactive decay. This leads to interesting processes resulting from such energy input, in terms of heating and thermal radiation.

The observation of a presence of radioactive material directly traces previous nucleosynthesis. It therefore can be used to locate cosmic nucleosynthesis events, and to measure their occurrence rates, depending if the radioactive decay time of a specific isotope is short, or long, respectively. For isotopes with a radioactive decay time that is longer than the typical other radiative signatures of nucleosynthesis events such as stellar explosions, measuring the $\gamma$~rays from radioactivity allows to trace the flow of freshly-produced nucleosynthesis ejecta over millions of years, along their path to mix with material that may form a later generation of stars. This allows for estimations of recycling times in the cosmic cycle of matter that leads to enrichment of metals in cosmic times, also called cosmic (or galactic) chemical evolution. 
Radioactivity in and following supernova explosions and their $\gamma$~rays also provide a new and different window into the processes that occur to launch and drive such explosions. In particular, relating $\gamma$-ray light curves and spectra to other observables characterising the supernova explosions can unfold dynamical processes in inner supernova regions, that otherwise are occulted from direct observations.  

\begin{acknowledgement}
R.D. acknowledges support from the Deutsche Forschungsgemeinschaft (DFG, German Research Foundation) under its Excellence Strategy, the Munich Clusters of Excellence \emph{Origin and Structure of the Universe} and \emph{Origins} (EXC-2094-390783311), by the EU through COST action ChETEC CA160117.
\end{acknowledgement}
%

\begin{thebibliography}{100}
\providecommand{\url}[1]{{#1}}
\providecommand{\urlprefix}{URL }
\expandafter\ifx\csname urlstyle\endcsname\relax
  \providecommand{\doi}[1]{DOI \discretionary{}{}{}#1}\else
  \providecommand{\doi}{DOI \discretionary{}{}{}\begingroup
  \urlstyle{rm}\Url}\fi

\bibitem{Patrignani:2016}
C.~{Patrignani}, e.~{(Particle Data Group)}, {Review of Particle Physics},
  Chinese Physics C \textbf{40}(100001) (2016)

\bibitem{Langanke:2003a}
K.~{Langanke}, G.~{Mart{\'{\i}}nez-Pinedo}, {Nuclear weak-interaction processes
  in stars}, Reviews of Modern Physics \textbf{75}, 819 (2003).
\newblock \doi{10.1103/RevModPhys.75.819}

\bibitem{Lichtenberg:2019}
T.~{Lichtenberg}, G.J. {Golabek}, R.~{Burn}, M.~{Meyer}, Y.~{Alibert},
  T.~{Gerya}, C.~{Mordasini}, in \emph{AAS/Division for Extreme Solar Systems
  Abstracts}, \emph{AAS/Division for Extreme Solar Systems Abstracts}, vol.~51
  (2019), \emph{AAS/Division for Extreme Solar Systems Abstracts}, vol.~51, p.
  311.01

\bibitem{Cyburt:2016}
R.H. {Cyburt}, B.D. {Fields}, K.A. {Olive}, T.H. {Yeh}, {Big bang
  nucleosynthesis: Present status}, Reviews of Modern Physics \textbf{88}(1),
  015004 (2016).
\newblock \doi{10.1103/RevModPhys.88.015004}

\bibitem{Asplund:2009}
M.~{Asplund}, N.~{Grevesse}, A.J. {Sauval}, P.~{Scott}, {The Chemical
  Composition of the Sun}, \araa \textbf{47}, 481 (2009).
\newblock \doi{10.1146/annurev.astro.46.060407.145222}

\bibitem{Merrill:1952}
P.W. {Merrill}, {Spectroscopic Observations of Stars of Class S}, \apj
  \textbf{116}, 21 (1952).
\newblock \doi{10.1086/145589}

\bibitem{Clayton:2018}
D.D. {Clayton}, in \emph{Astrophysics and Space Science Library},
  \emph{Astrophysics and Space Science Library}, vol. 453, ed. by R.~{Diehl},
  D.H. {Hartmann}, N.~{Prantzos} (2018), \emph{Astrophysics and Space Science
  Library}, vol. 453, p.~29.
\newblock \doi{10.1007/978-3-319-91929-4\_2}

\bibitem{Mewaldt:2001}
R.A. {Mewaldt}, N.E. {Yanasak}, M.E. {Wiedenbeck}, A.J. {Davis}, W.R. {Binns},
  E.R. {Christian}, A.C. {Cummings}, P.L. {Hink}, R.A. {Leske}, S.M. {Niebur},
  E.C. {Stone}, T.T. {Von Rosenvinge}, {Radioactive Clocks and Cosmic-ray
  Transport in the Galaxy}, Space Science Reviews \textbf{99}, 27 (2001)

\bibitem{Kutschera:2013}
W.~{Kutschera}, in \emph{European Physical Journal Web of Conferences},
  \emph{European Physical Journal Web of Conferences}, vol.~63 (2013),
  \emph{European Physical Journal Web of Conferences}, vol.~63, p. 03001.
\newblock \doi{10.1051/epjconf/20136303001}

\bibitem{Eddington:1919}
A.S. {Eddington}, {The sources of stellar energy}, The Observatory \textbf{42},
  371 (1919)

\bibitem{Arnett:1996a}
D.~{Arnett}, \emph{{Supernovae and nucleosynthesis. An investigation of the
  history of matter, from the Big Bang to the present}}.
\newblock Princeton series in astrophysics (Princeton University Press, New
  Jersey, USA) (1996)

\bibitem{Arnett:1989a}
W.D. {Arnett}, J.N. {Bahcall}, R.P. {Kirshner}, S.E. {Woosley}, {Supernova
  1987A}, \araa \textbf{27}, 629 (1989).
\newblock \doi{10.1146/annurev.aa.27.090189.003213}

\bibitem{Fransson:2002}
C.~{Fransson}, C.~{Kozma}, {Radioactivities and nucleosynthesis in SN 1987A},
  New Astronomy Review \textbf{46}, 487 (2002).
\newblock \doi{10.1016/S1387-6473(02)00188-4}

\bibitem{McCray:2016}
R.~{McCray}, C.~{Fransson}, {The Remnant of Supernova 1987A}, \araa
  \textbf{54}, 19 (2016).
\newblock \doi{10.1146/annurev-astro-082615-105405}

\bibitem{Scalzo:2014}
R.~{Scalzo}, G.~{Aldering}, P.~{Antilogus}, C.~{Aragon}, S.~{Bailey},
{\it et al.}, 
 {Nearby Supernova Factory},
  {Type Ia supernova bolometric light curves and ejected mass estimates from
  the Nearby Supernova Factory}, \mnras \textbf{440}(2), 1498 (2014).
\newblock \doi{10.1093/mnras/stu350}

\bibitem{Seitenzahl:2017}
I.R. {Seitenzahl}, D.M. {Townsley}, \emph{{Nucleosynthesis in Thermonuclear
  Supernovae}} (Springer: Berlin, Heidelberg),  p. 1955 (2017).
\newblock \doi{10.1007/978-3-319-21846-5.87}

\bibitem{Simnett:1981}
G.M. {Simnett}, in \emph{International Cosmic Ray Conference},
  \emph{International Cosmic Ray Conference}, vol.~12, pp. 205--227 (1981) 

\bibitem{Ryan:1979}
J.M. {Ryan}, L.L. {Chupp}, D.J. {Forrest}, M.L. {Cherry}, I.U. {Gleske},
{\it et al.}, 
in \emph{International Cosmic Ray
  Conference}, \emph{International Cosmic Ray Conference}, Vol.~5, p. 135 (1979)

\bibitem{Matz:1988}
S.M. {Matz}, G.H. {Share}, M.D. {Leising}, E.L. {Chupp}, W.T. {Vestrand},
  {Gamma-ray line emission from SN1987A}, \nat \textbf{331}, 416 (1988).
\newblock \doi{10.1038/331416a0}

\bibitem{Pluschke:2001c}
S.~{Pl{\"u}schke}, R.~{Diehl}, V.~{Sch{\"o}nfelder}, H.~{Bloemen},
  W.~{Hermsen}, {\it et al.}, 
 in \emph{Exploring the Gamma-Ray
  Universe}, \emph{ESA Special Publication}, vol. 459, pp. 55--58, ed. by A.~{Gimenez},
  V.~{Reglero}, C.~{Winkler} (2001)
  
\bibitem{Mahoney:1982}
W.A. {Mahoney}, J.C. {Ling}, A.S. {Jacobson}, R.E. {Lingenfelter}, {Diffuse
  galactic gamma-ray line emission from nucleosynthetic Fe-60, Al-26, and Na-22
  - Preliminary limits from HEAO 3}, \apj \textbf{262}, 742 (1982).
\newblock \doi{10.1086/160469}

\bibitem{Iben:1983}
J.~{Iben}, I., A.~{Renzini}, {Asymptotic giant branch evolution and beyond.},
  \araa \textbf{21}, 271 (1983).
\newblock \doi{10.1146/annurev.aa.21.090183.001415}

\bibitem{Meynet:1994b}
G.~{Meynet}, {Observation of the 1.8 MeV Emission Line from 26Al and the
  Chemical Evolution of the Galaxy}, \apjs \textbf{92}, 441 (1994).
\newblock \doi{10.1086/191992}

\bibitem{Clayton:1974b}
D.D. {Clayton}, F.~{Hoyle}, {Gamma-Ray Lines from Novae}, \apjl \textbf{187},
  L101 (1974).
\newblock \doi{10.1086/181406}

\bibitem{Timmes:1995}
F.X. {Timmes}, S.E. {Woosley}, D.H. {Hartmann}, R.D. {Hoffman}, T.A. {Weaver},
  F.~{Matteucci}, {26Al and 60Fe from Supernova Explosions}, \apj \textbf{449},
  204 (1995).
\newblock \doi{10.1086/176046}

\bibitem{Kozlovsky:1987}
B.~{Kozlovsky}, R.E. {Lingenfelter}, R.~{Ramaty}, {Positrons from accelerated
  particle interactions}, \apj \textbf{316}, 801 (1987).
\newblock \doi{10.1086/165244}

\bibitem{Schoenfelder:1993}
V.~{Sch\"onfelder}, H.~{Aarts}, K.~{Bennett}, H.~{de Boer}, J.~{Clear},
 {\it et al.}, 
{Instrument
  description and performance of the Imaging Gamma-Ray Telescope COMPTEL aboard
  the Compton Gamma-Ray Observatory}, \apjs \textbf{86}, 657 (1993).
\newblock \doi{10.1086/191794}

\bibitem{Gehrels:1993a}
N.~{Gehrels}, E.~{Chipman}, D.A. {Kniffen}, {The Compton Gamma Ray
  Observatory}, \aaps \textbf{97}, 5 (1993)

\bibitem{Diehl:1995}
R.~{Diehl}, {Imaging Diffuse Emission with COMPTEL}, Experimental Astronomy
  \textbf{6}, 103 (1995).
\newblock \doi{10.1007/BF00419264}

\bibitem{Diehl:1995b}
R.~{Diehl}, C.~{Dupraz}, K.~{Bennett}, H.~{Bloemen}, W.~{Hermsen},
{\it et al.}, 
  {COMPTEL observations of Galactic $^{26}$Al emission.}, \aap \textbf{298},
  445 (1995)

\bibitem{Knodlseder:1999}
J.~{Kn{\"o}dlseder}, D.~{Dixon}, K.~{Bennett}, H.~{Bloemen}, R.~{Diehl},
  W.~{Hermsen}, U.~{Oberlack}, J.~{Ryan}, V.~{Sch{\"o}nfelder}, P.~{von
  Ballmoos}, {Image reconstruction of COMPTEL 1.8 MeV (26) Al line data}, \aap
  \textbf{345}, 813 (1999)

\bibitem{Prantzos:1996a}
N.~{Prantzos}, R.~{Diehl}, {Radioactive 26Al in the galaxy: observations versus
  theory}, \physrep \textbf{267}, 1 (1996).
\newblock \doi{10.1016/0370-1573(95)00055-0}

\bibitem{Winkler:2003}
C.~{Winkler}, T.J.L. {Courvoisier}, G.~{Di Cocco}, N.~{Gehrels},
  A.~{Gim{\'e}nez}, {\it et al.}, 
{The INTEGRAL mission}, \aap \textbf{411}, L1
  (2003).
\newblock \doi{10.1051/0004-6361:20031288}

\bibitem{Vedrenne:2003}
G.~{Vedrenne}, J.P. {Roques}, V.~{Sch{\"o}nfelder}, P.~{Mandrou}, G.G.
  {Lichti}, {\it et al.}, 
  {SPI: The spectrometer aboard
  INTEGRAL}, \aap \textbf{411}, L63 (2003).
\newblock \doi{10.1051/0004-6361:20031482}

\bibitem{Diehl:2006d}
R.~{Diehl}, H.~{Halloin}, K.~{Kretschmer}, G.G. {Lichti}, V.~{Sch{\"o}nfelder},
{\it et al.}, 
 {Radioactive $^{26}$Al from massive stars in the
  Galaxy}, \nat \textbf{439}, 45 (2006).
\newblock \doi{10.1038/nature04364}

\bibitem{Kretschmer:2013}
K.~{Kretschmer}, R.~{Diehl}, M.~{Krause}, A.~{Burkert}, K.~{Fierlinger},
  O.~{Gerhard}, J.~{Greiner}, W.~{Wang}, {Kinematics of massive star ejecta in
  the Milky Way as traced by $^{26}$Al}, \aap \textbf{559}, A99 (2013).
\newblock \doi{10.1051/0004-6361/201322563}

\bibitem{Lingenfelter:1978}
R.E. {Lingenfelter}, R.~{Ramaty}, {Gamma-ray lines - A new window to the
  Universe}, Physics Today \textbf{31}, 40 (1978)

\bibitem{Krumholz:2018}
M.R. {Krumholz}, B.~{Burkhart}, J.C. {Forbes}, R.M. {Crocker}, {A unified model
  for galactic discs: star formation, turbulence driving, and mass transport},
  \mnras \textbf{477}(2), 2716 (2018).
\newblock \doi{10.1093/mnras/sty852}

\bibitem{Koo:2020}
B.C. {Koo}, C.G. {Kim}, S.~{Park}, E.C. {Ostriker}, {Radiative Supernova
  Remnants and Supernova Feedback}, \apj \textbf{905}(1), 35 (2020).
\newblock \doi{10.3847/1538-4357/abc1e7}

\bibitem{Diehl:2018f}
R.~{Diehl}, D.H. {Hartmann}, N.~{Prantzos}, \emph{{Distributed
  Radioactivities}} (Springer: Berlin, Heidelberg, vol. 453, chap.~7,
  pp. 427--500  (2018).
\newblock \doi{10.1007/978-3-319-91929-4_7}

\bibitem{Pleintinger:2020}
M.M.M. {Pleintinger}, Star groups and their nucleosynthesis.
\newblock Ph.D. thesis, Technische Universit{\"a}t M{\"u}nchen (2020)

\bibitem{Green:2019}
D.A. {Green}, {A revised catalogue of 294 Galactic supernova remnants}, Journal
  of Astrophysics and Astronomy \textbf{40}(4), 36 (2019).
\newblock \doi{10.1007/s12036-019-9601-6}

\bibitem{Chomiuk:2011}
L.~{Chomiuk}, M.S. {Povich}, {Toward a Unification of Star Formation Rate
  Determinations in the Milky Way and Other Galaxies}, \aj \textbf{142}, 197
  (2011).
\newblock \doi{10.1088/0004-6256/142/6/197}

\bibitem{Ahmad:2006}
I.~{Ahmad}, J.P. {Greene}, E.F. {Moore}, S.~{Ghelberg}, A.~{Ofan}, M.~{Paul},
  W.~{Kutschera}, {Improved measurement of the Ti44 half-life from a 14-year
  long study}, \prc \textbf{74}(6), 065803 (2006).
\newblock \doi{10.1103/PhysRevC.74.065803}

\bibitem{The:2006}
L.S. {The}, D.D. {Clayton}, R.~{Diehl}, D.H. {Hartmann}, A.F. {Iyudin}, M.D.
  {Leising}, B.S. {Meyer}, Y.~{Motizuki}, V.~{Sch{\"o}nfelder}, {Are
  $^{44}$Ti-producing supernovae exceptional?}, \aap \textbf{450}, 1037 (2006).
\newblock \doi{10.1051/0004-6361:20054626}

\bibitem{Dufour:2013}
F.~{Dufour}, V.M. {Kaspi}, {Limits on the Number of Galactic Young Supernova
  Remnants Emitting in the Decay Lines of $^{44}$Ti}, \apj \textbf{775}, 52
  (2013).
\newblock \doi{10.1088/0004-637X/775/1/52}

\bibitem{Dupraz:1997}
C.~{Dupraz}, H.~{Bloemen}, K.~{Bennett}, R.~{Diehl}, W.~{Hermsen}, A.F.
  {Iyudin}, J.~{Ryan}, V.~{Schoenfelder}, {COMPTEL three-year search for
  galactic sources of \^{}44\^{}Ti gamma-ray line emission at 1.157MeV.}, \aap
  \textbf{324}, 683 (1997)

\bibitem{Renaud:2006}
M.~{Renaud}, J.~{Vink}, A.~{Decourchelle}, F.~{Lebrun}, R.~{Terrier},
  J.~{Ballet}, {An INTEGRAL/IBIS view of young Galactic SNRs through the
  $^{44}$Ti gamma-ray lines}, New Astronomy Review \textbf{50}, 540 (2006).
\newblock \doi{10.1016/j.newar.2006.06.061}

\bibitem{Tsygankov:2016}
S.S. {Tsygankov}, R.A. {Krivonos}, A.A. {Lutovinov}, M.G. {Revnivtsev}, E.M.
  {Churazov}, R.A. {Sunyaev}, S.A. {Grebenev}, {Galactic survey of $^{44}$Ti
  sources with the IBIS telescope onboard INTEGRAL}, \mnras \textbf{458}, 3411
  (2016).
\newblock \doi{10.1093/mnras/stw549}

\bibitem{Weinberger:2020}
C.~{Weinberger}, R.~{Diehl}, M.M.M. {Pleintinger}, T.~{Siegert}, J.~{Greiner},
  {$^{44}$Ti ejecta in young supernova remnants}, \aap \textbf{638}, A83
  (2020).
\newblock \doi{10.1051/0004-6361/202037536}

\bibitem{Iyudin:1998}
A.F. {Iyudin}, V.~{Sch{\"o}nfelder}, K.~{Bennett}, H.~{Bloemen}, R.~{Diehl},
  W.~{Hermsen}, G.G. {Lichti}, R.D. {van der Meulen}, J.~{Ryan}, C.~{Winkler},
  {Emission from $^{44}$Ti associated with a previously unknown Galactic
  supernova}, \nat \textbf{396}, 142 (1998).
\newblock \doi{10.1038/24106}

\bibitem{Schonfelder:2000b}
V.~{Sch\"onfelder}, H.~{Bloemen}, W.~{Collmar}, R.~{Diehl}, W.~{Hermsen},
{\it et al.}, 
in \emph{American Institute of Physics Conference
  Series}, \emph{American Institute of Physics Conference Series}, vol. 510, pp. 54, 
  ed. by M.L. {McConnell}, J.M. {Ryan} (2000)
\newblock \doi{10.1063/1.1303173}

\bibitem{Slane:2001}
P.~{Slane}, J.P. {Hughes}, R.J. {Edgar}, P.P. {Plucinsky}, E.~{Miyata},
  H.~{Tsunemi}, B.~{Aschenbach}, {RX J0852.0-4622: Another Nonthermal
  Shell-Type Supernova Remnant (G266.2-1.2)}, \apj \textbf{548}, 814 (2001).
\newblock \doi{10.1086/319033}

\bibitem{Lin:2002}
R.P. {Lin}, B.R. {Dennis}, G.J. {Hurford}, D.M. {Smith}, A.~{Zehnder}, {\it et al.}, 
{The Reuven
  Ramaty High-Energy Solar Spectroscopic Imager (RHESSI)}, \solphys
  \textbf{210}, 3 (2002).
\newblock \doi{10.1023/A:1022428818870}

\bibitem{Smith:2003}
D.M. {Smith}, {The Reuven Ramaty High Energy Solar Spectroscopic Imager
  Observation of the 1809 keV Line from Galactic $^{26}$Al}, \apjl
  \textbf{589}, L55 (2003).
\newblock \doi{10.1086/375795}

\bibitem{Harris:2005}
M.J. {Harris}, J.~{Kn{\"o}dlseder}, P.~{Jean}, E.~{Cisana}, R.~{Diehl}, G.G.
  {Lichti}, J.P. {Roques}, S.~{Schanne}, G.~{Weidenspointner}, {Detection of
  {$\gamma$}-ray lines from interstellar $^{60}$Fe by the high resolution
  spectrometer SPI}, \aap \textbf{433}, L49 (2005).
\newblock \doi{10.1051/0004-6361:200500093}

\bibitem{Wang:2007}
W.~Wang, Study of long-lived radioactive sources in the galaxy with
  integral/spi.
\newblock Ph d thesis, TU Munich, Munich, Germany (2007)

\bibitem{Rauscher:2002}
T.~{Rauscher}, A.~{Heger}, R.D. {Hoffman}, S.E. {Woosley}, {Nucleosynthesis in
  Massive Stars with Improved Nuclear and Stellar Physics}, \apj \textbf{576},
  323 (2002).
\newblock \doi{10.1086/341728}

\bibitem{Woosley:2007b}
S.E. {Woosley}, A.~{Heger}, {Nucleosynthesis and remnants in massive stars of
  solar metallicity}, \physrep \textbf{442}, 269 (2007).
\newblock \doi{10.1016/j.physrep.2007.02.009}

\bibitem{Chieffi:2002}
A.~{Chieffi}, M.~{Limongi}, {The production of $^{26}$Al, $^{60}$Fe and
  $^{44}$Ti in massive stars of solar metallicity}, New Astronomy Review
  \textbf{46}, 459 (2002).
\newblock \doi{10.1016/S1387-6473(02)00183-5}

\bibitem{Limongi:2006c}
M.~{Limongi}, A.~{Chieffi}, {Nucleosynthesis of $^{60}$Fe in massive stars},
  \nar \textbf{50}, 474 (2006).
\newblock \doi{10.1016/j.newar.2006.06.005}

\bibitem{Limongi:2018}
M.~{Limongi}, A.~{Chieffi}, {Presupernova Evolution and Explosive
  Nucleosynthesis of Rotating Massive Stars in the Metallicity Range -3
  {\ensuremath{\leq}} [Fe/H] {\ensuremath{\leq}} 0}, \apjs \textbf{237}(1), 13
  (2018).
\newblock \doi{10.3847/1538-4365/aacb24}

\bibitem{Woosley:1997}
S.E. {Woosley}, {Neutron-rich Nucleosynthesis in Carbon Deflagration
  Supernovae}, \apj \textbf{476}, 801 (1997).
\newblock \doi{10.1086/303650}

\bibitem{Wang:2020}
W.~{Wang}, T.~{Siegert}, Z.G. {Dai}, R.~{Diehl}, J.~{Greiner}, A.~{Heger},
  M.~{Krause}, M.~{Lang}, M.M.M. {Pleintinger}, X.L. {Zhang}, {Gamma-Ray
  Emission of $^{60}$Fe and $^{26}$Al Radioactivity in Our Galaxy}, \apj
  \textbf{889}(2), 169 (2020).
\newblock \doi{10.3847/1538-4357/ab6336}

\bibitem{Wallner:2021}
A.~{Wallner}, M.B. {Froehlich}, M.A.C. {Hotchkis}, N.~{Kinoshita}, M.~{Paul},
{\it et al.}, 
{$^{60}$Fe and $^{244}$Pu
  deposited on Earth constrain the r-process yields of recent nearby
  supernovae}, Science \textbf{372}(6543), 742 (2021).
\newblock \doi{10.1126/science.aax3972}

\bibitem{Ellis:1996}
J.~{Ellis}, B.D. {Fields}, D.N. {Schramm}, {Geological Isotope Anomalies as
  Signatures of Nearby Supernovae}, \apj \textbf{470}, 1227 (1996).
\newblock \doi{10.1086/177945}

\bibitem{Wallner:2016}
A.~{Wallner}, J.~{Feige}, N.~{Kinoshita}, M.~{Paul}, L.K. {Fifield},
{\it et al.}, 
{Recent
  near-Earth supernovae probed by global deposition of interstellar radioactive
  $^{60}$Fe}, \nat \textbf{532}, 69 (2016).
\newblock \doi{10.1038/nature17196}

\bibitem{Breitschwerdt:2016}
D.~{Breitschwerdt}, J.~{Feige}, M.M. {Schulreich}, M.A.D. {Avillez},
  C.~{Dettbarn}, B.~{Fuchs}, {The locations of recent supernovae near the Sun
  from modelling $^{60}$Fe transport}, \nat \textbf{532}, 73 (2016).
\newblock \doi{10.1038/nature17424}

\bibitem{Clayton:1988b}
D.D. {Clayton}, {Meteorites: Studies of Nucleosynthesis}, \nat \textbf{332},
  683 (1988).
\newblock \doi{10.1038/332683a0}

\bibitem{Clayton:2004}
D.D. {Clayton}, L.R. {Nittler}, {Astrophysics with Presolar Stardust}, \araa
  \textbf{42}, 39 (2004).
\newblock \doi{10.1146/annurev.astro.42.053102.134022}

\bibitem{Zinner:2008}
E.~{Zinner}, {Stardust in the Laboratory}, \pasa \textbf{25}, 7 (2008).
\newblock \doi{10.1071/AS07039}

\bibitem{Israel:2018}
M.H. {Israel}, K.A. {Lave}, M.E. {Wiedenbeck}, W.R. {Binns}, E.R. {Christian},
{\it et al.}, 
 {Elemental Composition at the Cosmic-Ray
  Source Derived from the ACE-CRIS Instrument. I. $_{6}$C to $_{28}$Ni}, \apj
  \textbf{865}(1), 69 (2018).
\newblock \doi{10.3847/1538-4357/aad867}

\bibitem{Vink:2012}
J.~{Vink}, {Supernova remnants: the X-ray perspective}, \aapr \textbf{20}, 49
  (2012).
\newblock \doi{10.1007/s00159-011-0049-1}

\bibitem{Reynolds:2008}
S.P. {Reynolds}, {Supernova remnants at high energy.}, \araa \textbf{46}, 89
  (2008).
\newblock \doi{10.1146/annurev.astro.46.060407.145237}

\bibitem{Tueller:1988}
J.~{Tueller}, S.~{Barthelmy}, N.~{Gehrels}, B.J. {Teegarden}, M.~{Leventhal},
  C.J. {MacCallum}, in \emph{Nuclear Spectroscopy of Astrophysical Sources},
  \emph{American Institute of Physics Conference Series}, vol. 170, pp. 439--443, ed. by
  N.~{Gehrels}, G.H. {Share} (1988)
  \newblock \doi{10.1063/1.37243}

\bibitem{Naya:1996}
J.E. {Naya}, S.D. {Barthelmy}, L.M. {Bartlett}, N.~{Gehrels}, M.~{Leventhal},
  A.~{Parsons}, B.J. {Teegarden}, J.~{Tueller}, {Detection of high-velocity
  $^{26}$Al towards the Galactic Centre}, \nat \textbf{384}, 44 (1996).
\newblock \doi{10.1038/384044a0}

\bibitem{Chen:1997}
W.~{Chen}, R.~{Diehl}, N.~{Gehrels}, D.~{Hartmann}, M.~{Leising}, J.E. {Naya},
  N.~{Prantzos}, J.~{Tueller}, P.~{von Ballmoos}, in \emph{The Transparent
  Universe}, \emph{ESA Special Publication}, vol. 382, pp. 105, ed. by C.~{Winkler},
  T.J.L. {Courvoisier}, P.~{Durouchoux} (1997)
  
\bibitem{Sturner:1999}
S.J. {Sturner}, J.E. {Naya}, {On the Nature of the High-Velocity $^{26}$Al near
  the Galactic Center}, \apj \textbf{526}, 200 (1999).
\newblock \doi{10.1086/307979}

\bibitem{Diehl:2018}
R.~{Diehl}, T.~{Siegert}, J.~{Greiner}, M.~{Krause}, K.~{Kretschmer},
 {\it et al.}, 
  {INTEGRAL/SPI {$\gamma$}-ray line spectroscopy. Response and background
  characteristics}, \aap \textbf{611}, A12 (2018).
\newblock \doi{10.1051/0004-6361/201731815}

\bibitem{Krause:2015}
M.G.H. {Krause}, R.~{Diehl}, Y.~{Bagetakos}, E.~{Brinks}, A.~{Burkert},
  O.~{Gerhard}, J.~{Greiner}, K.~{Kretschmer}, T.~{Siegert}, {$^{26}$Al
  kinematics: superbubbles following the spiral arms?. Constraints from the
  statistics of star clusters and HI supershells}, \aap \textbf{578}, A113
  (2015).
\newblock \doi{10.1051/0004-6361/201525847}

\bibitem{Krause:2021}
M.G.H. {Krause}, D.~{Rodgers-Lee}, J.E. {Dale}, R.~{Diehl}, C.~{Kobayashi},
  {Galactic $^{26}$Al traces metal loss through hot chimneys}, \mnras
  \textbf{501}(1), 210 (2021).
\newblock \doi{10.1093/mnras/staa3612}

\bibitem{Schinnerer:2019}
E.~{Schinnerer}, A.~{Hughes}, A.~{Leroy}, B.~{Groves}, G.A. {Blanc},
{\it et al.}, 
 {The Gas-Star Formation Cycle
  in Nearby Star-forming Galaxies. I. Assessment of Multi-scale Variations},
  \apj \textbf{887}(1), 49 (2019).
\newblock \doi{10.3847/1538-4357/ab50c2}

\bibitem{Rodgers-Lee:2019}
D.~{Rodgers-Lee}, M.G.H. {Krause}, J.~{Dale}, R.~{Diehl}, {Synthetic $^{26}$Al
  emission from galactic-scale superbubble simulations}, \mnras
  \textbf{490}(2), 1894 (2019).
\newblock \doi{10.1093/mnras/stz2708}

\bibitem{Nath:2020}
B.B. {Nath}, P.~{Das}, M.S. {Oey}, {Size distribution of superbubbles}, \mnras
  \textbf{493}(1), 1034 (2020).
\newblock \doi{10.1093/mnras/staa336}

\bibitem{Krause:2020}
M.G.H. {Krause}, S.S.R. {Offner}, C.~{Charbonnel}, M.~{Gieles}, R.S. {Klessen},
{\it et al.}, 
{The Physics of Star
  Cluster Formation and Evolution}, \ssr \textbf{216}(4), 64 (2020).
\newblock \doi{10.1007/s11214-020-00689-4}

\bibitem{Chevance:2022}
M.~{Chevance}, J.M.D. {Kruijssen}, M.R. {Krumholz}, B.~{Groves}, B.W. {Keller},
 {\it et al.}, 
{Pre-supernova feedback mechanisms drive the
  destruction of molecular clouds in nearby star-forming disc galaxies}, \mnras
  \textbf{509}(1), 272 (2022).
\newblock \doi{10.1093/mnras/stab2938}

\bibitem{Burrows:1993}
D.N. {Burrows}, K.P. {Singh}, J.A. {Nousek}, G.P. {Garmire}, J.~{Good}, {A
  multiwavelength study of the Eridanus soft X-ray enhancement}, \apj
  \textbf{406}, 97 (1993).
\newblock \doi{10.1086/172423}

\bibitem{Heiles:1999}
C.~{Heiles}, L.M. {Haffner}, R.J. {Reynolds}, in \emph{New Perspectives on the
  Interstellar Medium}, \emph{Astronomical Society of the Pacific Conference
  Series}, vol. 168, pp. 211, ed. by {A.~R.~Taylor, T.~L.~Landecker, \& G.~Joncas}
  (1999)

\bibitem{Siegert:2017a}
T.~{Siegert}, R.~{Diehl}, in \emph{14th International Symposium on Nuclei in
  the Cosmos (NIC2016)}, p. 020305, ed. by S.~{Kubono}, T.~{Kajino}, S.~{Nishimura},
  T.~{Isobe}, S.~{Nagataki}, T.~{Shima}, Y.~{Takeda} (2017)
\newblock \doi{10.7566/JPSCP.14.020305}

\bibitem{Fierlinger:2016}
K.M. {Fierlinger}, A.~{Burkert}, E.~{Ntormousi}, P.~{Fierlinger},
  M.~{Schartmann}, A.~{Ballone}, M.G.H. {Krause}, R.~{Diehl}, {Stellar feedback
  efficiencies: supernovae versus stellar winds}, \mnras \textbf{456}, 710
  (2016).
\newblock \doi{10.1093/mnras/stv2699}

\bibitem{Alexis:2014}
A.~{Alexis}, P.~{Jean}, P.~{Martin}, K.~{Ferri{\`e}re}, {Monte Carlo modelling
  of the propagation and annihilation of nucleosynthesis positrons in the
  Galaxy}, \aap \textbf{564}, A108 (2014).
\newblock \doi{10.1051/0004-6361/201322393}

\bibitem{Knodlseder:2005}
J.~{Kn{\"o}dlseder}, P.~{Jean}, V.~{Lonjou}, G.~{Weidenspointner},
  N.~{Guessoum}, {\it et al.}, 
 {The all-sky distribution of 511 keV electron-positron
  annihilation emission}, \aap \textbf{441}, 513 (2005).
\newblock \doi{10.1051/0004-6361:20042063}

\bibitem{Jean:2006}
P.~{Jean}, J.~{Kn{\"o}dlseder}, W.~{Gillard}, N.~{Guessoum}, K.~{Ferri{\`e}re},
  A.~{Marcowith}, V.~{Lonjou}, J.P. {Roques}, {Spectral analysis of the
  Galactic positron annihilation emission}, \aap \textbf{445}, 579 (2006).
\newblock \doi{10.1051/0004-6361:20053765}

\bibitem{Siegert:2016}
T.~{Siegert}, R.~{Diehl}, G.~{Khachatryan}, M.G.H. {Krause}, F.~{Guglielmetti},
  J.~{Greiner}, A.W. {Strong}, X.~{Zhang}, {Gamma-ray spectroscopy of positron
  annihilation in the Milky Way}, \aap \textbf{586}, A84 (2016).
\newblock \doi{10.1051/0004-6361/201527510}

\bibitem{Churazov:2020}
E.~{Churazov}, L.~{Bouchet}, P.~{Jean}, E.~{Jourdain}, J.~{Kn{\"o}dlseder},
{\it et al.}, 
{INTEGRAL results on the electron-positron annihilation
  radiation and X-ray \& Gamma-ray diffuse emission of the Milky Way}, \nar
  \textbf{90}, 101548 (2020).
\newblock \doi{10.1016/j.newar.2020.101548}

\bibitem{Martin:2010}
P.~{Martin}, J.~{Vink}, S.~{Jiraskova}, P.~{Jean}, R.~{Diehl}, {Annihilation
  emission from young supernova remnants}, \aap \textbf{519}, A100 (2010).
\newblock \doi{10.1051/0004-6361/201014171}

\bibitem{Prantzos:2011}
N.~{Prantzos}, C.~{Boehm}, A.M. {Bykov}, R.~{Diehl}, K.~{Ferri{\`e}re},
 {\it et al.}, 
{The 511 keV emission from
  positron annihilation in the Galaxy}, Reviews of Modern Physics \textbf{83},
  1001 (2011).
\newblock \doi{10.1103/RevModPhys.83.1001}

\bibitem{Arnett:1982}
W.D. {Arnett}, {Type I supernovae. I - Analytic solutions for the early part of
  the light curve}, \apj \textbf{253}, 785 (1982).
\newblock \doi{10.1086/159681}

\bibitem{Brachwitz:2000}
F.~{Brachwitz}, D.J. {Dean}, W.R. {Hix}, K.~{Iwamoto}, K.~{Langanke},
  G.~{Mart{\'\i}nez-Pinedo}, K.~{Nomoto}, M.R. {Strayer}, F.K. {Thielemann},
  H.~{Umeda}, {The Role of Electron Captures in Chandrasekhar-Mass Models for
  Type IA Supernovae}, \apj \textbf{536}(2), 934 (2000).
\newblock \doi{10.1086/308968}

\bibitem{Mori:2018}
K.~{Mori}, M.A. {Famiano}, T.~{Kajino}, T.~{Suzuki}, P.M. {Garnavich}, G.J.
  {Mathews}, R.~{Diehl}, S.C. {Leung}, K.~{Nomoto}, {Nucleosynthesis
  Constraints on the Explosion Mechanism for Type Ia Supernovae}, \apj
  \textbf{863}(2), 176 (2018).
\newblock \doi{10.3847/1538-4357/aad233}

\bibitem{Yamaguchi:2014}
H.~{Yamaguchi}, C.~{Badenes}, R.~{Petre}, T.~{Nakano}, D.~{Castro}, {\it et al.}, 
{Discriminating the Progenitor Type of
  Supernova Remnants with Iron K-shell Emission}, \apjl \textbf{785}(2), L27
  (2014).
\newblock \doi{10.1088/2041-8205/785/2/L27}

\bibitem{Colgate:1969}
S.A. {Colgate}, C.~{McKee}, {Early Supernova Luminosity}, \apj \textbf{157},
  623 (1969).
\newblock \doi{10.1086/150102}

\bibitem{Clayton:1974}
D.D. {Clayton}, S.E. {Woosley}, {Thermonuclear astrophysics}, Reviews of Modern
  Physics \textbf{46}, 755 (1974).
\newblock \doi{10.1103/RevModPhys.46.755}

\bibitem{Churazov:2014}
E.~{Churazov}, R.~{Sunyaev}, J.~{Isern}, J.~{Kn{\"o}dlseder}, P.~{Jean},
 {\it et al.}, 
{Cobalt-56 {$\gamma$}-ray emission lines from the type Ia
  supernova 2014J}, \nat \textbf{512}, 406 (2014).
\newblock \doi{10.1038/nature13672}

\bibitem{Fossey:2014aa}
J.~{Fossey}, B.~{Cooke}, G.~{Pollack}, M.~{Wilde}, T.~{Wright}, {Supernova
  2014J in M82 = Psn J09554214+6940260}, Central Bureau Electronic Telegrams
  \textbf{3792}, 1 (2014)

\bibitem{Foley:2014}
R.J. {Foley}, O.D. {Fox}, C.~{McCully}, M.M. {Phillips}, D.J. {Sand},
{\it et al.}, 
  {Extensive HST ultraviolet spectra and multiwavelength observations of SN
  2014J in M82 indicate reddening and circumstellar scattering by typical
  dust}, \mnras \textbf{443}, 2887 (2014).
\newblock \doi{10.1093/mnras/stu1378}

\bibitem{Diehl:2014}
R.~{Diehl}, T.~{Siegert}, W.~{Hillebrandt}, S.A. {Grebenev}, J.~{Greiner},
{\it et al.}, 
  {Early $^{56}$Ni decay gamma rays from SN2014J suggest an unusual explosion},
  Science \textbf{345}(6201), 1162 (2014).
\newblock \doi{10.1126/science.1254738}

\bibitem{Diehl:2015}
R.~{Diehl}, T.~{Siegert}, W.~{Hillebrandt}, M.~{Krause}, J.~{Greiner},
  K.~{Maeda}, F.K. {R{\"o}pke}, S.A. {Sim}, W.~{Wang}, X.~{Zhang}, {SN2014J
  gamma rays from the $^{56}$Ni decay chain}, \aap \textbf{574}, A72 (2015).
\newblock \doi{10.1051/0004-6361/201424991}

\bibitem{Isern:2016}
J.~{Isern}, P.~{Jean}, E.~{Bravo}, J.~{Kn{\"o}dlseder}, F.~{Lebrun},
{\it et al.}, 
{Gamma-ray emission from SN2014J near maximum optical
  light}, \aap \textbf{588}, A67 (2016).
\newblock \doi{10.1051/0004-6361/201526941}

\bibitem{Summa:2013}
A.~{Summa}, A.~{Ulyanov}, M.~{Kromer}, S.~{Boyer}, F.K. {R{\"o}pke}, {\it et al.}, 
  {Gamma-ray diagnostics of Type Ia supernovae. Predictions of observables from
  three-dimensional modeling}, \aap \textbf{554}, A67 (2013).
\newblock \doi{10.1051/0004-6361/201220972}

\bibitem{Dhawan:2016}
S.~{Dhawan}, B.~{Leibundgut}, J.~{Spyromilio}, S.~{Blondin}, {A reddening-free
  method to estimate the $^{56}$Ni mass of Type Ia supernovae}, \aap
  \textbf{588}, A84 (2016).
\newblock \doi{10.1051/0004-6361/201527201}

\bibitem{The:2014}
L.S. {The}, A.~{Burrows}, {Expectations for the Hard X-Ray Continuum and
  Gamma-Ray Line Fluxes from the Type Ia Supernova SN 2014J in M82}, \apj
  \textbf{786}, 141 (2014).
\newblock \doi{10.1088/0004-637X/786/2/141}

\bibitem{Leising:1990}
M.D. {Leising}, G.H. {Share}, {The gamma-ray light curves of SN 1987A}, \apj
  \textbf{357}, 638 (1990).
\newblock \doi{10.1086/168952}

\bibitem{Tueller:1990}
J.~{Tueller}, S.~{Barthelmy}, N.~{Gehrels}, B.J. {Teegarden}, M.~{Leventhal},
  C.J. {MacCallum}, {Observations of gamma-ray line profiles from SN 1987A},
  \apjl \textbf{351}, L41 (1990).
\newblock \doi{10.1086/185675}

\bibitem{Jerkstrand:2020}
A.~{Jerkstrand}, A.~{Wongwathanarat}, H.T. {Janka}, M.~{Gabler}, D.~{Alp},
 {\it et al.}, 
  {Properties of gamma-ray decay lines in 3D core-collapse supernova models,
  with application to SN 1987A and Cas A}, \mnras \textbf{494}(2), 2471 (2020).
\newblock \doi{10.1093/mnras/staa736}

\bibitem{Seitenzahl:2014}
I.R. {Seitenzahl}, F.X. {Timmes}, G.~{Magkotsios}, {The Light Curve of SN 1987A
  Revisited: Constraining Production Masses of Radioactive Nuclides}, \apj
  \textbf{792}(1), 10 (2014).
\newblock \doi{10.1088/0004-637X/792/1/10}

\bibitem{Grebenev:2012}
S.A. {Grebenev}, A.A. {Lutovinov}, S.~{Tsygankov}, C.~{Winkler}, Hard-x-ray
  emission lines from the decay of $^{44}$ti in the remnant of supernova 1987a,
  Nature \textbf{(tbd)}((tbd)), (accepted for publication) (2012)

\bibitem{Boggs:2015}
S.E. {Boggs}, F.A. {Harrison}, H.~{Miyasaka}, B.W. {Grefenstette},
  A.~{Zoglauer}, {\it et al.}, 
 {$^{44}$Ti gamma-ray
  emission lines from SN1987A reveal an asymmetric explosion}, Science
  \textbf{348}, 670 (2015).
\newblock \doi{10.1126/science.aaa2259}

\bibitem{Harrison:2013}
F.A. {Harrison}, W.W. {Craig}, F.E. {Christensen}, C.J. {Hailey}, W.W. {Zhang},
 {\it et al.}, 
 {The Nuclear
  Spectroscopic Telescope Array (NuSTAR) High-energy X-Ray Mission}, \apj
  \textbf{770}, 103 (2013).
\newblock \doi{10.1088/0004-637X/770/2/103}

\bibitem{Woosley:1991}
S.E. {Woosley}, R.D. {Hoffman}, {57Co and 44Ti Production in SN 1987A}, \apjl
  \textbf{368}, L31 (1991).
\newblock \doi{10.1086/185941}

\bibitem{Timmes:1996a}
F.X. {Timmes}, S.E. {Woosley}, D.H. {Hartmann}, R.D. {Hoffman}, {The Production
  of 44Ti and 60Co in Supernovae}, \apj \textbf{464}, 332 (1996).
\newblock \doi{10.1086/177323}

\bibitem{Nagataki:1998}
S.~{Nagataki}, M.A. {Hashimoto}, K.~{Sato}, S.~{Yamada}, Y.S. {Mochizuki}, {The
  High Ratio of 44Ti/ 56Ni in Cassiopeia A and the Axisymmetric Collapse-driven
  Supernova Explosion}, \apjl \textbf{492}, L45 (1998).
\newblock \doi{10.1086/311089}

\bibitem{Magkotsios:2008}
G.~{Magkotsios}, F.X. {Timmes}, M.~{Wiescher}, C.L. {Fryer}, A.~{Hungerford},
 {\it et al.}, 
in \emph{Proceedings of the 10th Symposium
  on Nuclei in the Cosmos (NIC X). July 27 - August 1, 2008 Mackinac Island,
  Michigan, USA. Available online at
  http://pos.sissa.it/cgi-bin/reader/conf.cgi?confid=53} (2008)

\bibitem{Magkotsios:2010}
G.~{Magkotsios}, F.X. {Timmes}, A.L. {Hungerford}, C.L. {Fryer}, P.A. {Young},
  M.~{Wiescher}, {Trends in $^{44}$Ti and $^{56}$Ni from Core-collapse
  Supernovae}, \apjs \textbf{191}, 66 (2010).
\newblock \doi{10.1088/0067-0049/191/1/66}

\bibitem{Sukhbold:2016}
T.~{Sukhbold}, T.~{Ertl}, S.E. {Woosley}, J.M. {Brown}, H.T. {Janka},
  {Core-collapse Supernovae from 9 to 120 Solar Masses Based on
  Neutrino-powered Explosions}, \apj \textbf{821}, 38 (2016).
\newblock \doi{10.3847/0004-637X/821/1/38}

\bibitem{Wongwathanarat:2017}
A.~{Wongwathanarat}, H.T. {Janka}, E.~{M{\"u}ller}, E.~{Pllumbi}, S.~{Wanajo},
  {Production and Distribution of $^{44}$Ti and $^{56}$Ni in a
  Three-dimensional Supernova Model Resembling Cassiopeia A}, \apj
  \textbf{842}(1), 13 (2017).
\newblock \doi{10.3847/1538-4357/aa72de}

\bibitem{Curtis:2019}
S.~{Curtis}, K.~{Ebinger}, C.~{Fr{\"o}hlich}, M.~{Hempel}, A.~{Perego},
  M.~{Liebend{\"o}rfer}, F.K. {Thielemann}, {PUSHing Core-collapse Supernovae
  to Explosions in Spherical Symmetry. III. Nucleosynthesis Yields}, \apj
  \textbf{870}(1), 2 (2019).
\newblock \doi{10.3847/1538-4357/aae7d2}

\bibitem{Fesen:2006}
R.A. {Fesen}, M.C. {Hammell}, J.~{Morse}, R.A. {Chevalier}, K.J. {Borkowski},
{\it et al.}, 
 {The Expansion Asymmetry and Age of the Cassiopeia A Supernova
  Remnant}, \apj \textbf{645}(1), 283 (2006).
\newblock \doi{10.1086/504254}

\bibitem{Grefenstette:2014}
B.W. {Grefenstette}, F.A. {Harrison}, S.E. {Boggs}, S.P. {Reynolds}, C.L.
  {Fryer}, {\it et al.}, 
 {Asymmetries in core-collapse supernovae from
  maps of radioactive $^{44}$Ti in CassiopeiaA}, \nat \textbf{506}, 339 (2014).
\newblock \doi{10.1038/nature12997}

\bibitem{Iyudin:1994}
A.F. {Iyudin}, R.~{Diehl}, H.~{Bloemen}, W.~{Hermsen}, G.G. {Lichti},
 {\it et al.}, 
 {COMPTEL observations of Ti-44 gamma-ray line
  emission from CAS A}, \aap \textbf{284}, L1 (1994)

\bibitem{The:1996}
L.S. {The}, M.D. {Leising}, J.D. {Kurfess}, W.N. {Johnson}, D.H. {Hartmann},
  N.~{Gehrels}, J.E. {Grove}, W.R. {Purcell}, {CGRO/OSSE observations of the
  Cassiopeia A SNR.}, \aaps \textbf{120}, C357+ (1996)

\bibitem{Rothschild:1999}
R.E. {Rothschild}, R.E. {Lingenfelter}, P.R. {Blanco}, D.E. {Gruber}, W.A.
  {Heindl}, {\it et al.}, 
{RXTE
  observations of Cas A}, Nuclear Physics B Proceedings Supplements
  \textbf{69}, 68 (1999).
\newblock \doi{10.1016/S0920-5632(98)00186-8}

\bibitem{Vink:2000}
J.~{Vink}, J.S. {Kaastra}, J.A.M. {Bleeker}, H.~{Bloemen}, {The Hard X-Ray
  Emission and 44ti Emission Of Cas A}, Advances in Space Research \textbf{25},
  689 (2000).
\newblock \doi{10.1016/S0273-1177(99)00823-6}

\bibitem{Siegert:2015}
T.~{Siegert}, R.~{Diehl}, M.G.H. {Krause}, J.~{Greiner}, {Revisiting
  INTEGRAL/SPI observations of $^{44}$Ti from Cassiopeia A}, \aap \textbf{579},
  A124 (2015).
\newblock \doi{10.1051/0004-6361/201525877}

\bibitem{Grefenstette:2017}
B.W. {Grefenstette}, C.L. {Fryer}, F.A. {Harrison}, S.E. {Boggs}, T.~{DeLaney},
{\it et al.}, 
{The Distribution of Radioactive $^{44}$Ti in
  Cassiopeia A}, \apj \textbf{834}, 19 (2017).
\newblock \doi{10.3847/1538-4357/834/1/19}

\bibitem{Weinberger:2021}
C.~{Weinberger}, Supernova diagnostics from gamma-ray lines in the young
  remnant phase.
\newblock Ph.D. thesis, TU Munich (2021)

\bibitem{Seitenzahl:2015}
I.R. {Seitenzahl}, A.~{Summa}, F.~{Krau{\ss}}, S.A. {Sim}, R.~{Diehl},
{\it et al.}, 
  {5.9-keV Mn K-shell X-ray luminosity from the decay of $^{55}$Fe in Type Ia
  supernova models}, \mnras \textbf{447}(2), 1484 (2015).
\newblock \doi{10.1093/mnras/stu2537}

\bibitem{Borkowski:2010}
K.J. {Borkowski}, S.P. {Reynolds}, D.A. {Green}, U.~{Hwang}, R.~{Petre},
{\it et al.}, 
  {Radioactive Scandium in the Youngest
  Galactic Supernova Remnant G1.9+0.3}, \apjl \textbf{724}, L161 (2010).
\newblock \doi{10.1088/2041-8205/724/2/L161}

\end{thebibliography}

\end{document}